\documentclass[11pt]{article}
\usepackage[margin=1.05in]{geometry}
\usepackage[T1]{fontenc}
\usepackage{lmodern,microtype,amsmath,amssymb,amsthm,mathtools,bm}
\usepackage{booktabs,longtable,array,enumitem}
\usepackage{xcolor}
\usepackage{tikz}
\usetikzlibrary{arrows.meta,positioning}
\usepackage[backend=bibtex,style=alphabetic,natbib=true,sorting=nyt,backref=true,
 maxcitenames=4,mincitenames=1,maxalphanames=5,minalphanames=3,maxbibnames=99]{biblatex}
\renewbibmacro*{pageref}{\printlist[pageref][-\value{listtotal}]{pageref}}

\usepackage[colorlinks,linkcolor=blue!55!black,citecolor=blue!55!black,urlcolor=blue!55!black]{hyperref}
\usepackage[nameinlink,noabbrev]{cleveref}
\newtheorem{theorem}{Theorem}[section]
\newtheorem{lemma}[theorem]{Lemma}
\newtheorem{proposition}[theorem]{Proposition}

\theoremstyle{definition}
\newtheorem{definition}[theorem]{Definition}

\newtheorem{remark}[theorem]{Remark}

\crefname{algo}{Algorithm}{Algorithms}
\DeclareMathOperator{\MC}{MAXCUT}
\DeclareMathOperator{\Var}{Var}
\DeclareMathOperator{\Cov}{Cov}

\DeclareMathOperator{\KL}{D_{KL}}
\DeclareMathOperator{\TV}{TV}
\DeclareMathOperator{\diag}{diag}

\newcommand{\E}{\mathbb E}
\newcommand{\R}{\mathbb R}
\newcommand{\one}{\boldsymbol 1}
\newcommand{\eps}{\varepsilon}
\newcommand{\norm}[1]{\left\lVert#1\right\rVert}
\newcommand{\ip}[2]{\left\langle#1,#2\right\rangle}

\newcommand{\Unif}{\mathsf U}

\newcommand{\mcald}{\mathcal D}

\newcommand{\cutg}{{\operatorname{cut}_G}}

\allowdisplaybreaks[2]
\title{Testing MAX-CUT in the Adjacency Matrix Model:\\Tight Non-Adaptive Bounds and a Cubic Adaptive Lower Bound}
\author{Songhua He\\Rutgers University\\songhuahe.cs@gmail.com}
\date{October 2026}
\begin{document}
\maketitle
\begin{abstract}

We study the query complexity of approximating the MAX-CUT value to additive error $\varepsilon n^2$ in the adjacency matrix model. This was one of the first problems studied in graph property testing \cite{GoldreichGoldwasserRon1998}.
Previously, the best known upper bound for both adaptive and non-adaptive algorithms was $O(1/\varepsilon^6)$ \cite{RudelsonVershynin2007,MathieuSchudy2008,GhoshMishraRaychaudhurySen2022}. The only known lower bound was a trivial $\Omega(1/\varepsilon^2)$ bound.
We close this gap from both sides for non-adaptive algorithms, proving that its query complexity is exactly $\Theta(1/\varepsilon^4)$. We also prove an adaptive $\Omega(1/\varepsilon^3)$ lower bound.

Our upper bound is achieved by a canonical tester that samples $O(1/\varepsilon^2)$ vertices in expectation and outputs the rescaled MAX-CUT value of the induced subgraph. We prove its correctness by applying spectral truncation to the adjacency matrix, using a geometric argument to control the sampling error of the retained low-rank part and a cut-norm restriction estimate \cite{RudelsonVershynin2007} for the remainder. As an application, an analogous proof also yields a tight $O(1/\eps^4)$ non-adaptive query upper bound for MAX-2CSP.

For our lower bounds, we show indistinguishability between a uniform random graph and a random graph with a planted hidden partition. For the non-adaptive lower bound, we observe that one plus the $\chi^2$-divergence between the query transcript distributions is exactly a normalized Ising partition function, which can be bounded using Boolean hypercontractivity.
For the adaptive lower bound, we relate each query's distinguishing power to the prediction, based on previous answers, of whether its endpoints lie in the same part of the hidden partition. We bound the combined advantage of these predictions using the completed transcript, which lets us analyze the adaptively chosen pairs of vertices together despite their dependence on earlier answers.

\end{abstract}

\clearpage
\section{Introduction}\label{sec:intro}
Property testing asks whether a gigantic combinatorial object has a
given property, or is far from having it, while inspecting only a tiny
fraction of the input. In the \emph{adjacency matrix model}, an $n$-vertex
graph is given by oracle access to its adjacency matrix, and the
distance $\eps$ of a graph from a property is the fraction of
adjacency-matrix entries that must be flipped to satisfy the property.
This setting is natural when the graph encodes a dense binary relation,
such as bipartiteness~\cite{GoldreichGoldwasserRon1998,alon2002testing,BogdanovTrevisan2002},
MAX-CUT~\cite{GoldreichGoldwasserRon1998,AlonFernandezKannanKarpinski2003,MathieuSchudy2008,GhoshMishraRaychaudhurySen2022}, or subgraph-freeness~\cite{alon2002testing1,alon2006characterization,AlonShapira2008}.
Many natural
properties admit testers whose query complexity depends only on $\eps$
and not on $n$; the celebrated work of~\citet{AlonShapira2008} characterized the properties admitting oblivious one-sided-error testers with query complexity depending only on $\varepsilon$ as exactly the semi-hereditary properties.

Among testable properties, the \emph{easily testable} ones---those
admitting testers of query complexity polynomial in $1/\eps$---have
attracted much interest. 
While MAX-CUT is a canonical NP-hard and even APX-hard graph problem
\cite{PapadimitriouYannakakis1991}, estimating the MAX-CUT value is a
different task: here we study its query complexity rather than running
time.

Estimating the MAX-CUT value is among the
first problems shown to be easily testable: given query access to
the adjacency matrix of an $n$-vertex graph $G$, output a value within
additive error $\eps n^2$ of $\MC(G)$. This also yields an estimate of the distance from bipartiteness:
in the normalization above, that distance is
exactly $2(|E(G)|-\MC(G))/n^2$, where $|E(G)|$ can be estimated within additive error $\eps n^2$ in $O(1/\eps^2)$ queries. This
estimation is the approximation counterpart of bipartiteness testing in
\citet{GoldreichGoldwasserRon1998}. 

The seminal work of \citet{GoldreichGoldwasserRon1998} gave the first tester for this problem, with $\widetilde O(1/\varepsilon^7)$ non-adaptive queries. 
Subsequent work focused on the simpler class of \emph{canonical testers}, which sample a uniformly random set of vertices and query the induced subgraph: a line of work by \citet{AlonFernandezKannanKarpinski2003,RudelsonVershynin2007} and independent work by \citet{MathieuSchudy2008} established an $O(1/\eps^8)$ non-adaptive query upper bound using $O(1/\eps^4)$ sampled vertices. \citet{GhoshMishraRaychaudhurySen2022} recorded a folklore improvement of the non-adaptive upper bound to $O(1/\varepsilon^6)$ achieved by only querying a subset of pairs from the induced subgraph. On the lower bound side, there is only a trivial $\Omega(1/\eps^2)$ lower bound.

\Cref{tab:previous-upper-bounds} compares the number of vertices sampled\footnote{In some works, the number of vertices sampled by an algorithm is also called its \emph{sample complexity}.} and the query complexity of each algorithm.
\begin{table}[htbp]
\centering
\begin{tabular}{@{}p{0.39\linewidth}p{0.24\linewidth}p{0.29\linewidth}@{}}
\toprule
Work & Sampled vertices & Query complexity \\
\midrule
\cite{GoldreichGoldwasserRon1998}
 & $O(\log(1/\eps)/\eps^5)$ & $O(\log^2(1/\eps)/\eps^7)$ \\
\cite{AlonFernandezKannanKarpinski2003}
 & $O(\log(1/\eps)/\eps^4)$ & $O(\log^2(1/\eps)/\eps^8)$ \\
\cite{RudelsonVershynin2007}
 & $O(1/\eps^4)$ & $O(1/\eps^8)$ \\
\cite{MathieuSchudy2008}
 & $O(1/\eps^4)$ & $O(1/\eps^8)$ \\
\cite{MathieuSchudy2008,GhoshMishraRaychaudhurySen2022}
 & $O(1/\eps^4)$ & $O(1/\eps^6)$ \\
This work & $O(1/\eps^2)$ & $O(1/\eps^4)$ \\
\bottomrule
\end{tabular}
\caption{Upper bounds for estimating the MAX-CUT value.}
\label{tab:previous-upper-bounds}
\end{table}

We settle the non-adaptive query complexity of this problem, giving both an $O(1/\varepsilon^4)$ non-adaptive upper bound and an $\Omega(1/\eps^4)$ non-adaptive lower bound. We also give an $\Omega(1/\eps^3)$ adaptive lower bound for this problem.\footnote{ChatGPT was used to generate the technical proofs in this paper. The author subsequently substantially rewrote and verified them; see the Use of AI statement in the Acknowledgements on page~\pageref{sec:use-of-ai}.}

Our main result is an optimal non-adaptive query upper bound for MAX-CUT.

\begin{theorem}[Informal statement for \Cref{thm:upper}]
    There exists a non-adaptive algorithm of query complexity $O(1/\varepsilon^4)$ that approximates the MAX-CUT value to within additive error $\varepsilon n^2$ with probability $\ge 2/3$.
\end{theorem}

Our algorithm is also a canonical tester for MAX-CUT \cite{AlonFernandezKannanKarpinski2003,RudelsonVershynin2007,MathieuSchudy2008}: we query the induced subgraph of a uniformly sampled vertex set. The output is a rescaled estimate of the MAX-CUT value of the induced subgraph. We prove that $\Theta(1/\eps^2)$ vertices suffice to give an accurate estimate.
\citet{GoldreichTrevisan2003} pointed out that any property that is testable with $q$ adaptive queries can also be tested by canonical testers on a sample set of size $O(q)$.
In this work, we show that a canonical tester achieves the optimal non-adaptive query complexity for MAX-CUT.

Efficient property testing algorithms for MAX-CUT have also been proposed by \citet{GhoshMishraRaychaudhurySen2022}. They studied MAX-CUT in the tolerant testing setting: they gave $\tilde O(1/\eps^3)$ testing algorithms for distinguishing whether a graph is at most $\varepsilon$-close-to-bipartite or at least $((2+c)\varepsilon)$-far-from-bipartite, for $c>0$. Their algorithm does not generalize to estimating the MAX-CUT value of arbitrary graphs.

Our analysis is in spirit closer to that of \cite{AlonFernandezKannanKarpinski2003,RudelsonVershynin2007}. While they analyze the cut-norm decomposition of the adjacency matrix \cite{FriezeKannan1999}, we apply a spectral truncation of the adjacency matrix, decomposing it
into a low-rank part and a remainder. We reduce controlling the sampling
error of the low-rank part to bounding the Hausdorff distance between
two carefully constructed convex hulls, one for the full graph and one
for the sampled graph. We control the remainder by a cut-norm estimate
\cite{RudelsonVershynin2007}. As a result, we
show that the MAX-CUT value of a uniformly random set of $O(1/\eps^2)$ vertices suffices to estimate the MAX-CUT value of the whole graph.

An analogous proof also yields a tight $O(1/\eps^4)$ non-adaptive query upper bound for MAX-2CSP. An instance consists of $n$ Boolean variables and a set of distinct Boolean constraints, each specified on a pair of variables; a pair may carry multiple constraints. Querying a pair reveals all its constraints. The goal is to estimate the maximum number of constraints satisfied by an assignment to within additive error $\eps n^2$.

\begin{theorem}[Informal statement for \Cref{thm:csp-upper}]
There exists a non-adaptive algorithm that makes $O(1/\eps^4)$ queries and estimates the MAX-2CSP value to within additive error $\eps n^2$ with probability at least $2/3$.
\end{theorem}

\citet{AlonFernandezKannanKarpinski2003} showed that sampling $O(\log(1/\eps)/\eps^4)$ variables and solving the induced instance gives an estimate after rescaling. \citet{RudelsonVershynin2007} and independently \citet{MathieuSchudy2008} reduced the number of sampled variables to $O(1/\eps^4)$. Our argument shows that $O(1/\eps^2)$ sampled variables suffice.

MAX-CUT is a special case of MAX-2CSP: place the constraint $x_u\ne x_v$ on each edge $(u,v)$ and no constraint on nonedges. Thus both of our lower bounds for MAX-CUT also apply to MAX-2CSP. The upper bound \Cref{thm:csp-upper} is proved in Appendix \ref{app:max2csp}.

To complement the above upper bound, we give a matching non-adaptive lower bound for this problem.

\begin{theorem}[Informal statement for \Cref{thm:na}]
    For sufficiently small $\varepsilon$ and $n\ge \lceil 1/\eps^2\rceil$,
    any possibly randomized non-adaptive algorithm that approximates the MAX-CUT value within additive error $\varepsilon n^2$ with probability $\ge 2/3$ must have worst-case query complexity $\Omega(1/\varepsilon^{4})$.
\end{theorem}

We achieve it by showing the indistinguishability of a pair of input distributions. The no distribution $\mcald_0$ is the uniform graph, in
which each pair is an edge independently with probability $1/2$; the yes
distribution $\mcald_1$ plants a uniformly random hidden partition $\Sigma\in\{\pm 1\}^n$, with each $\Sigma_u\in\{\pm1\}$ denoting the label of the part containing vertex $u$, and connects each pair of vertices $(u,v)$ with independent probability $\frac{1-\rho\Sigma_u\Sigma_v}{2}$, for $\rho=\Theta(\eps)$. So crossing
pairs become slightly more likely and within-part pairs slightly less.
The two distributions differ in MAX-CUT value by $\Theta(\eps n^2)$, yet any fixed pair of vertices is an edge with probability $1/2$ under both distributions. Thus an algorithm must exploit correlations among the answers. These hard distributions are natural. Similar distributions have already been used to lower-bound MAX-CUT and related
problems in other models, such as query lower bounds for MAX-CUT in the adjacency list model \cite{ChiplunkarEtAl2018}, and the time-space trade-offs in the random-query model \cite{GargHePapakonstantinou2026}.

Prior to our work, there was no non-trivial lower bound for MAX-CUT known in the literature, even in the non-adaptive setting. There was only a simple $\Omega(1/\varepsilon^2)$ query lower bound, which follows from the same hard distributions we use: even if an algorithm knows the hidden partition in advance, the probability that a queried pair is an edge differs between the yes and no distributions by only $\Theta(\varepsilon)$. Each pair query therefore reveals at most 
$$\KL(\textnormal{Bern}(1/2)\|\textnormal{Bern}(1/2\pm \Theta(\varepsilon)))=\Theta(\varepsilon^2)$$ 
bits of information, so distinguishing the two distributions requires $\Omega(1/\varepsilon^2)$ queries.

Our lower bound follows from a simple observation. Fix a deterministic and non-adaptive algorithm of query complexity $q$. The algorithm will always query the same set of pairs of vertices. We denote by $H$ the graph that contains exactly these pairs of vertices. One plus the $\chi^2$-divergence between the transcript distributions under the yes and no inputs is exactly a normalized Ising partition function on $H$. We subsequently bound it using Boolean hypercontractivity \cite{ODonnell2014}. We note that we do not use Ising-specific tools to obtain this bound. Nevertheless, this is a clean connection between the Ising model and our hard distributions.

This connection also has a combinatorial interpretation.
The distributions of query answers on $H$ under the yes and no inputs differ if and only if $H$ contains a cycle.
Along any cycle of queried pairs, multiplying the
hidden signs $\Sigma_u\Sigma_v$ gives $1$, since every vertex
appears twice. Since the answer to each pair query $(u,v)$\footnote{For convenience, we encode the answer to a pair query $(u,v)$ as $Y_{uv}=+1$ for a nonedge and $Y_{uv}=-1$ for an edge.} has mean $\rho\Sigma_u\Sigma_v$ under the yes
distribution, the product of the answers
on a cycle of length $\ell$ has mean $\rho^\ell$. In contrast, the product has mean zero under the no distribution.
More generally, the expectations of query answer products are nonzero precisely for
sets of queried pairs in which every vertex has even degree. The
$\chi^2$-divergence sums the squares of these products over nonempty sets.
The normalized Ising partition function has the same even-degree expansion, with the empty edge set contributing the additional $1$.
We will derive this identity more formally in \Cref{subsec:overview-na}.

Our connection to the Ising model differs from two others in the literature. One involves the hard distributions we use; the other involves MAX-CUT itself. \citet{AbbeBoix2020} observed that the conditional distribution of $\Sigma$ given the answers to a fixed set of queried pairs is an Ising model. Separately, MAX-CUT has a classical Ising formulation: given a graph $G$ with adjacency matrix $A$, taking $-A$ as the coupling matrix makes its minimum-energy configurations correspond exactly to maximum cuts of $G$ \citep{BarahonaGroetschelJungerReinelt1988,ZdeborovaBoettcher2010}.

Our last result is a new adaptive lower bound for MAX-CUT.

\begin{theorem}[Informal statement for \Cref{thm:cubic}]
    For sufficiently small $\varepsilon$ and $n\ge \lceil 1/\eps^2\rceil$,
    any possibly randomized adaptive algorithm that approximates the MAX-CUT value within additive error $\varepsilon n^2$ with success probability at least $99/100$\footnote{The same asymptotic lower bound also holds for success probability $2/3$ by standard amplification.} must make $\Omega(1/\varepsilon^3)$ pair queries in the worst case.
\end{theorem}

Since the query graph $H$ is no longer fixed on an adaptive algorithm, the observation above cannot be easily extended to an adaptive lower bound. We instead view its run as a sequence of implicit predictions: after choosing a pair $(u,v)$ and before seeing its answer, the algorithm's current transcript determines a conditional expectation of $\Sigma_u\Sigma_v$. Under the yes distribution, this prediction determines how far the next answer is biased away from a fair coin. The KL chain rule therefore bounds the divergence between the yes and no transcript distributions by $\rho^2$ times the expected sum of the squares of these predictions. Our key step is to replace each prediction based on the current history with the prediction based on the complete transcript. The complete transcript may weaken a particular prediction, but its squared magnitude can only increase on average. We can then study all queried pairs under a single conditional distribution of the hidden partition $\Sigma$, even though the pairs were chosen adaptively. We show that strong correlations across many queried pairs would require the transcript to reveal substantial information about the hidden partition. Bounding the first two moments of this information controls the expected correlation sum over all transcripts at once. This completes the cubic adaptive lower bound.

\paragraph{Organization.}

\Cref{sec:overview} gives the proof overview of all results.
\Cref{sec:model} introduces notation and preliminaries.
\Cref{sec:upper} proves the quartic non-adaptive query upper bound.
\Cref{sec:hard} defines the hard distributions that we use for proving lower bounds.
\Cref{sec:nonadaptive} proves the matching quartic non-adaptive lower bound.
\Cref{sec:warmup} proves the cubic adaptive lower bound.

\section{Proof overview}\label{sec:overview}
We first outline the quartic non-adaptive upper bound, then the matching non-adaptive lower bound and the cubic adaptive lower bound.

\subsection{The $O(1/\varepsilon^4)$ upper bound}

We give a proof overview of our quartic non-adaptive query upper bound of MAX-CUT. The full proof is given in \Cref{sec:upper}.

The algorithm is a canonical tester. Fix the input
graph $G$ and an expected sample size $1\le s\le n$.
We sample a set $Q$ obtained by including each vertex independently with probability $s/n$. The tester queries every pair in $Q$ and returns
$$
 \frac{n^2}{s^2}\MC(G[Q]).
$$
The normalization uses the expected size $s$. Our main conclusion is that the expected sampling error is
$$
 \E\left[\left|\frac{\MC(G[Q])}{s^2}
                  -\frac{\MC(G)}{n^2}\right|\right]=O(1/\sqrt s),
$$
which implies that a sample set $Q$ of expected size $s=\Theta(1/\eps^2)$ would suffice to achieve additive error $\eps n^2$ with constant probability.

Let $A$ be the adjacency matrix of $G$. We rewrite the MAX-CUT value as
$$
 \MC(G)=\frac14\left(\one^TA\one
       -\min_{\sigma\in\{-1,1\}^n}\sigma^TA\sigma\right).
$$
Writing $A_{QQ}$ for the adjacency matrix of $G[Q]$, the triangle
inequality gives, for every sample $Q$,
$$
 \begin{aligned}
 &\left|\frac{\MC(G[Q])}{s^2}-\frac{\MC(G)}{n^2}\right|\\
 \le&\frac14
 \underbrace{\left|\frac{\one^TA_{QQ}\one}{s^2}
                    -\frac{\one^TA\one}{n^2}\right|}
             _{\text{edge-density error}}+\frac14
 \underbrace{\left|
 \frac{\min_{\sigma\in\{-1,1\}^{Q}}\sigma^TA_{QQ}\sigma}{s^2}
 -\frac{\min_{\sigma\in\{-1,1\}^{n}}\sigma^TA\sigma}{n^2}
 \right|}_{\text{minimum-value error}}.
 \end{aligned}
$$
We bound the expectations of these two errors separately and then add them; no independence is needed. A variance bound for $\one^TA_{QQ}\one$, together with Cauchy--Schwarz, gives an expected edge-density error of at most $3/\sqrt s$. The hard part is bounding the minimum-value error.

A natural idea is to apply a concentration bound to show that the induced cut of every fixed cut on $Q$ is close to its expected value. However, this does not work since there are $2^n$ cuts, and a union bound over all of them would require a sample size that depends on $n$.

\paragraph{Earlier approaches.}

To overcome this obstruction, earlier work either replaces the $2^n$
cuts by a small structured family of candidates \cite{GoldreichGoldwasserRon1998, MathieuSchudy2008}, or uses a cut-norm
decomposition of the adjacency matrix into few rectangles \cite{AlonFernandezKannanKarpinski2003,RudelsonVershynin2007}. Our approach is
closest to the latter, but we use a spectral truncation of the adjacency
matrix instead of a rectangular cut decomposition.

\citet[Section~8.1]{GoldreichGoldwasserRon1998} divide $V$ into
$\ell=\Theta(1/\eps)$ blocks. For each block, they sample a small set
of $\tilde O(1/\eps^2)$ vertices outside it and try every bipartition of this seed; each bipartition of the seed induces a bipartition of the block by assigning each vertex to the side containing fewer of its neighbors in the seed.
These partitions combined together implicitly define $2^{\tilde O(1/\eps^3)}$ candidate
cuts. A hybrid argument that replaces the blocks of an optimal cut
one at a time shows that some candidate cut has size close to the optimum.
Since they output the maximum estimated cut value over all candidate cuts, the union bound ranges over candidates rather
than all $2^n$ cuts. This yields a $\tilde O(1/\eps^7)$ non-adaptive
upper bound.

\citet{MathieuSchudy2008} analyze random-order
greedy algorithms for MAX-CUT on dense graphs. This algorithm
enumerates all cuts of a small sample. Starting from each such cut, it processes the remaining vertices
in random order, assigning each vertex to the side that maximizes
the current cut value, and outputs the best completed cut. They show that this gives a PTAS, using
martingale concentration bounds in the analysis. Their analysis
also shows that the expected normalized MAX-CUT
value of an induced subgraph on $s=\Theta(1/\eps^4)$ random
vertices differs from the full-graph value only by $O(\eps)$. This directly gives an $O(1/\eps^8)$ non-adaptive query upper bound by querying all pairs in the sample.

\citet{AlonFernandezKannanKarpinski2003} use the
Frieze--Kannan decomposition \cite{FriezeKannan1999} to write the
adjacency matrix as a sum of $O(1/\eps^2)$ rectangles plus a remainder
small in cut norm. They show that a uniformly random set $Q$ of $s=\Theta(\log(1/\eps)/\eps^4)$ vertices gives $|(n/s)^2\MC(G[Q])-\MC(G)|\le\eps n^2$ with high probability. Their proof shows that the remainder
stays small in cut norm after random restriction, then uses a
linear-programming argument to compare the optimal values of the
rectangular approximation before and after sampling. \citet{RudelsonVershynin2007} sharpen the cut-norm restriction estimate used in this argument, removing the logarithmic factor from the required sample size.

\citet[Appendix~A]{GhoshMishraRaychaudhurySen2022} record a
folklore observation that the query complexity of canonical testers
such as those of
\cite{AlonFernandezKannanKarpinski2003, RudelsonVershynin2007, MathieuSchudy2008}
can be reduced by querying a random subset of vertex pairs within
the sampled set $Q$, rather than all pairs. On a sample $Q$ of $s$ vertices,
$O(s/\eps^2)$ pair queries suffice to estimate all cut values
simultaneously. Combining this observation with the
upper bound of
\citet{MathieuSchudy2008} gives
an $O(1/\eps^6)$ non-adaptive upper bound.

\paragraph{Our approach.}

We apply spectral truncation to $A$. For a threshold $0<\tau\le1$
to be chosen later, write
$$
 L=\sum_{|\theta_j|>\tau n}\theta_j u_ju_j^T,\qquad R=A-L,
$$
where $\theta_j$ and $u_j$ are the eigenvalues and orthonormal
eigenvectors of $A$. At most $1/\tau^2$ eigenvalues are retained in
$L$, while the remainder satisfies
$$
 \norm R_{\mathrm{op}}\le\tau n,\qquad \norm R_F\le\norm A_F\le n.
$$
The advantage of $L$ is that, for any cut vector
$x\in\{-1,1\}^n$, the value $x^TLx$ depends only on the projections
of $x$ onto the retained eigenvectors. This lets us express the
optimization problem for $L$ in at most $1/\tau^2$ dimensions,
independently of the number of vertices.

Let $P$ be the diagonal matrix with $P_{ii}=1$ for $i\in Q$ and
$P_{ii}=0$ otherwise. Thus $PAP$ agrees with $A_{QQ}$ on the sampled
vertices and is zero elsewhere. Since $A$ and $PAP$ have zero diagonal,
minimizing $x^TAx$ or $x^TPAPx$ over $\{-1,1\}^n$ is equivalent to
minimizing it over $[-1,1]^n$: with all other coordinates fixed,
the objective is linear in each coordinate. Below, $x$ ranges over
$[-1,1]^n$. We replace $A$ by $L$ in these two minimization problems
and account for the error introduced by discarding $R$. The triangle
inequality gives
\begin{align}
 \left|\frac{\min_x x^TPAPx}{s^2}-\frac{\min_x x^TAx}{n^2}\right|
 &\le\frac1{s^2}\sup_x|x^TPRPx|\notag\\
 &\quad+\left|\frac{\min_x x^TPLPx}{s^2}
                 -\frac{\min_x x^TLx}{n^2}\right|\notag\\
 &\quad+\frac1{n^2}\sup_x|x^TRx|.
\label{eq:overview-error-decomposition}
\end{align}
The three terms are the cost of discarding $R$ in the sample, the
sampling error for $L$, and the cost of discarding $R$ in the full
graph. The last is at most $\tau$, since
$|x^TRx|\le\norm R_{\mathrm{op}}\norm{x}_2^2\le\tau n^2$.
We now bound the first two terms.

For the sampled remainder, the same operator-norm argument gives
$$
 |x^TPRPx|=|(Px)^TR(Px)|
 \le\norm R_{\mathrm{op}}\norm{Px}_2^2
 \le\tau n|Q|,
$$
since $\norm{Px}_2^2=\sum_{i\in Q}x_i^2\le|Q|$ for every
$x\in[-1,1]^n$. Taking the supremum over $x$, then expectations,
and using $\E[|Q|]=s$, we obtain only
$$
 \frac1{s^2}\E\left[\sup_x|x^TPRPx|\right]
 \le\frac{\tau n}{s^2}\E[|Q|]=\frac{\tau n}{s}.
$$
To obtain a bound independent of $n$, we use the cut-norm estimate
of \citet{RudelsonVershynin2007}. Let the
$\ell_\infty\to\ell_1$ norm $\norm B_{\infty\to1}$ denote the maximum
of $|x^TBy|$ over $x,y\in[-1,1]^n$, so it bounds $|x^TBx|$ for every
$x$ simultaneously. Applying their estimate to $R$ gives
\begin{align}
 \frac1{s^2}\E\left[\sup_x|x^TPRPx|\right]
 &\le\frac{\E[\norm{PRP}_{\infty\to1}]}{s^2}\notag\\
 &=O\left(\tau+\frac1{\tau s}+\frac1{\sqrt s}\right).
\label{eq:overview-residual}
\end{align}
The norm and diagonal estimates needed
for this application are proved in \Cref{subsec:residual}.

The technically heavier part is the second term of \eqref{eq:overview-error-decomposition}, which is the sampling error for $L$. We next compare the two minima for $L$ by expressing them as minima
of the same function over two convex sets.
To see the intuition, we express the objectives by the spectral expansion of $L$:
\begin{align*}
 \frac{x^TLx}{n^2}
 &=\frac1{n^2}x^T(\sum_{j}\theta_ju_ju_j^T)x\\
 &=\frac1{n^2}\sum_j\theta_j\sum_{i,k}x_ix_ku_j(i)u_j(k)\\
 &=\frac1{n^2}\sum_j\theta_j\left(\sum_{i}x_iu_j(i)\right)^2,\\
 \frac{x^TPLPx}{s^2}&=\frac1{s^2}\sum_j\theta_j\left(\sum_{i\in Q}x_iu_j(i)\right)^2.
\end{align*}
We rescale the eigenvalues and eigenvectors by
setting $\lambda_j=\theta_j/n$ and $\phi_j(i)=\sqrt n\,u_j(i)$. For each vertex $i$, define $y_i=(\lambda_j\phi_j(i))_j$, and let
$f(w)=\sum_j w_j^2/\lambda_j$.
These definitions express both objectives as values of the same function $f$, evaluated at normalized weighted sums of the $y_i$ over all vertices and over $Q$, respectively:
\begin{align*}
 f\left(\frac1n\sum_i x_iy_i\right)&=\sum_j\frac{1}{\lambda_j}\left(\frac1n\sum_ix_i\lambda_j\phi_j(i)\right)^2\\
 &=\frac1{n^2}\sum_j\theta_j\left(\sum_ix_iu_j(i)\right)^2\\
 &=\frac{x^TLx}{n^2},\\
 f\left(\frac1s\sum_{i\in Q}x_iy_i\right)
 &=\frac1{s^2}\sum_j\theta_j
   \left(\sum_{i\in Q}x_iu_j(i)\right)^2\\
 &=\frac{x^TPLPx}{s^2}.
\end{align*}

As $x$ ranges over $[-1,1]^n$, the arguments of $f$ range over the
compact convex sets
$$
 K=\left\{\frac1n\sum_i x_iy_i:x\in[-1,1]^n\right\},\qquad
 K_Q=\left\{\frac1s\sum_{i\in Q}x_iy_i:x\in[-1,1]^n\right\}.
$$
The two minima for $L$ are therefore $\min_K f$ and $\min_{K_Q}f$.
Our plan is first to bound their Euclidean Hausdorff distance
$\Delta=d_H(K_Q,K)$, which guarantees that every point of either set
has a point of the other within distance $\Delta$. 
Then, we compare nearby points $w\in K$ and $z\in K_Q$ by expanding $f(z)$ around $w$. Since $f$ is quadratic, its Taylor expansion at $w\in K$ is exact and uses only the gradient at $w$ and the fixed quadratic coefficients $1/\lambda_j$. We bound the gradient on $K$ using orthogonality and the coefficients using $|\lambda_j|>\tau$; no gradient bound between $w$ and $z$ is needed. 

Geometrically, each $x_i\in[-1,1]$ selects a point on the line
segment from $-y_i$ to $y_i$. The set $K$ consists of all sums of
these points divided by $n$; $K_Q$ uses only the sampled segments
and divides by $s$. The eigenvalue weights ensure that the vectors
$y_i$ have bounded average squared length:
$$
 t_y:=\frac1n\sum_i\norm{y_i}_2^2
 =\sum_j\lambda_j^2\le\frac{\norm A_F^2}{n^2}\le1.
$$
By Cauchy--Schwarz, the contribution of any $\alpha n$ vertices to
a normalized sum defining $K$ has Euclidean norm at most
$\sqrt{\alpha t_y}\le\sqrt{\alpha}$. Thus no small subset of vertices can make a large contribution. In \Cref{lem:hausdorff-moments}, we bound the first two moments
of the Hausdorff distance between $K$ and $K_Q$ in terms of
$t_y$ and $s$, independently of the number of retained eigenvectors.
We use ghost-sample symmetrization and the Rademacher contraction
principle to bound the expectation, and the Efron--Stein inequality
to control the variance.

Using a gradient bound on $K$ and a second-order Taylor expansion
of $f$, we obtain
$$
 \left|\min_{K_Q}f-\min_K f\right|\le2\Delta+\Delta^2/\tau,
$$
where $\|\nabla f(w)\|_2\le2$ for every $w\in K$, and the coefficients of the quadratic term are $1/\lambda_j$, each of absolute value at most $1/\tau$.

Taking expectations and using the bounds on the first two moments
of $\Delta$ from \Cref{lem:hausdorff-moments}, we obtain
\begin{equation}
 \E\left[\left|\min_{K_Q}f-\min_K f\right|\right]
 =O\left(\frac1{\sqrt s}+\frac1{\tau s}\right).
\label{eq:overview-retained}
\end{equation}

Taking expectations in \eqref{eq:overview-error-decomposition},
applying \eqref{eq:overview-residual} and \eqref{eq:overview-retained},
and adding the $O(1/\sqrt s)$ edge-density error gives
$$
 \E\left[\left|\frac{\MC(G[Q])}{s^2}-\frac{\MC(G)}{n^2}\right|\right]
 =O\left(\tau+\frac1{\tau s}+\frac1{\sqrt s}\right).
$$
Choosing $\tau=1/\sqrt s$ makes this $O(1/\sqrt s)$. Taking
$s=\lceil C/\eps^2\rceil$ for a sufficiently large constant $C$,
Markov's inequality gives the required accuracy with constant success
probability. We obtain a MAX-CUT estimator that makes only $O(1/\varepsilon^4)$ non-adaptive queries.

\paragraph{MAX-2CSP.} MAX-2CSP is a natural generalization of MAX-CUT. The MAX-2CSP algorithm of \citet{AlonFernandezKannanKarpinski2003} uses the same canonical tester and overall proof strategy as the MAX-CUT case discussed above.

Specifically, fix an ordering of the two variables in each constraint and define four matrices $A^{(a,b)}, a,b\in\{0,1\}$. Each entry $A^{(a,b)}_{ij}$ counts constraints assigned to the ordered pair $(i,j)$ that are satisfied when $x_i=a$ and $x_j=b$.
For every assignment to $x$, its MAX-2CSP value is exactly
$$\sum_{a,b\in\{0,1\}}\sum_{i,j\in[n]}A_{ij}^{(a,b)}(1-x_i)^{1-a}x_i^a(1-x_j)^{1-b}x_j^b.$$
They applied Frieze--Kannan decomposition~\cite{FriezeKannan1999} to rewrite these matrices and analyzed the contribution of their random submatrices. The following work of \citet{RudelsonVershynin2007} refined their cut norm analysis and improved their query complexity to $O(1/\eps^8)$. Independently, \citet{MathieuSchudy2008} obtained the same $O(1/\eps^8)$ non-adaptive query upper bound for MAX-2CSP.

For our bound, we combine these four matrices into a symmetric matrix
$M$ indexed by variable-value pairs $(i,a)$. For $i\ne j$, set
$M_{(i,a),(j,b)}=(A_{ij}^{(a,b)}+A_{ji}^{(b,a)})/2$; for $i=j$, set
$M_{(i,a),(i,b)}=0$ for all $a,b\in\{0,1\}$. For an assignment $x$, set $z_{i,x_i}=1$ and $z_{i,1-x_i}=0$ for every $i$. Its MAX-2CSP value is then $z^TMz$.
We relax this optimization problem by allowing each $z_{i,a}$ to be fractional, over the following set
$$
 \mathcal X=\{z\in[0,1]^{[n]\times\{0,1\}}:
                 z_{i,0}+z_{i,1}=1\text{ for every }i\}.
$$
This does not change the maximum: with all other variables fixed,
$z^TMz$ is linear in both $z_{i,0}$ and $z_{i,1}$. Thus $(z_{i,0},z_{i,1})$ can be moved to an endpoint, $(1,0)$ or $(0,1)$, without decreasing the value. The same argument applies to
the sampled instance. As above, let $Q$ be the sampled set of variables.
Let $P$ be the $2n\times2n$ diagonal matrix whose entries at $(i,0)$ and
$(i,1)$ are both $1$ if $i\in Q$ and $0$ otherwise. Our goal is to compare
$\frac1{n^2}\max_{z\in\mathcal X}z^TMz$ and
$\frac1{s^2}\max_{z\in\mathcal X}z^TPMPz$.

We apply spectral truncation $M=L+R$ at the same threshold $\tau n$. As in \eqref{eq:overview-error-decomposition}, we decompose the error into three terms and bound each using the corresponding tool from the MAX-CUT analysis. We defer the details to \Cref{app:max2csp}.

\subsection{Hard distributions and a simple non-adaptive lower bound}\label{subsec:overview-na}
By Yao's minimax principle, to give query lower bounds for randomized algorithms, it suffices to give a pair of hard distributions and show that (i) their MAX-CUT values are separated by $\eps n^2$ with high probability, and (ii) any algorithm that makes fewer than the claimed number of queries cannot distinguish them with high probability.

We use the following pair of hard distributions throughout all of our lower bounds. Let $\rho=\Theta(\eps)$ be a large enough multiple of $\eps$. The \emph{no} distribution $\mcald_0$ is the uniform distribution on all graphs with $n$ vertices. The \emph{yes} distribution $\mcald_1$ generates a graph together with a hidden partition $\Sigma\in\{-1,1\}^n$ as follows. First, sample $\Sigma$ uniformly at random. Then, for each pair of vertices $(u,v)$, include the edge $(u,v)$ independently with probability $(1-\rho\Sigma_u\Sigma_v)/2$. We denote $Y_{uv}\in\{\pm1\}$ as its edge indicator, which evaluates to $-1$ if and only if $(u,v)$ are connected by an edge. By concentration bounds, the cut size of the hidden partition is at least $2\varepsilon n^2$ larger than the MAX-CUT value under the no distribution with high probability. We write $\mcald_1^{\mathrm{joint}}$ for the joint distribution of $\Sigma$ and the graph under $\mcald_1$.

\paragraph{The non-adaptive quartic lower bound.}

We start by sketching the proof of the non-adaptive $\Omega(1/\varepsilon^4)$ query lower bound. The key observation is that the $\chi^2$-divergence between the yes and no transcript distributions plus one is exactly a normalized Ising partition function on the query graph.

For any fixed deterministic non-adaptive algorithm of query complexity $q$, the algorithm always queries a fixed set of $q$ pairs from the input graph. We denote by $H$ the \emph{query graph} whose edges are exactly these $q$ queried pairs. Its edge set $E(H)$ records which pairs are queried, whether or not they are edges of the input graph $G$. Let $\mcald_{1,H},\mcald_{0,H}$ denote the distributions of the answer vector $(Y_{uv})_{(u,v)\in E(H)}$ under $\mcald_1,\mcald_0$, respectively. We have

\begin{align*}
 1+\chi^2(\mcald_{1,H}\Vert\mcald_{0,H})
 &=\E_{Y\sim\mcald_{0,H}}\left[
   \left(\frac{\mcald_{1,H}(Y)}{\mcald_{0,H}(Y)}\right)^2\right]\\
 &=\E_Y\left[\left(\E_\Sigma\left[
   \prod_{(u,v)\in E(H)}(1+\rho Y_{uv}\Sigma_u\Sigma_v)
   \right]\right)^2\right]\\
 &=\E_{\Sigma,\Sigma'}\left[
   \prod_{(u,v)\in E(H)}\E_{Y_{uv}}\left[
   (1+\rho Y_{uv}\Sigma_u\Sigma_v)
   (1+\rho Y_{uv}\Sigma'_u\Sigma'_v)\right]\right]\\
 &=\E_\Xi\left[
   \prod_{(u,v)\in E(H)}(1+\rho^2\Xi_u\Xi_v)\right].
\end{align*}
Here $\Sigma,\Sigma'$ are independent uniform hidden partitions,
independent of $Y\sim\mcald_{0,H}$, and
$\Xi_v:=\Sigma_v\Sigma'_v$ is also a uniformly random vector.
We use the facts that $\E[Y_{uv}]=0$ on the no distribution and $Y_{uv}^2=1$.

We then bound its exponential
moment by the Boolean hypercontractivity
\citep[Theorem~9.21]{ODonnell2014} bound when $q\rho^4$ is small, giving the non-adaptive quartic lower bound. Details of the proof are deferred to \Cref{sec:nonadaptive}.

As noted in the introduction, the Ising partition function above also has a combinatorial interpretation. Given that $\Xi$ is uniform on $\{\pm1\}^n$, only terms corresponding to sets of queried pairs in which every vertex has even degree survive under expectation.

\begin{align*}
 \E_\Xi\left[\prod_{(u,v)\in E(H)}(1+\rho^2\Xi_u\Xi_v)\right]
 &=\sum_{F\subseteq E(H)}\rho^{2|F|}
   \E_\Xi\left[\prod_{(u,v)\in F}\Xi_u\Xi_v\right]\\
 &=\sum_{\substack{F\subseteq E(H):\\
                  \text{every }v\in[n]\text{ has even degree in }F}}\rho^{2|F|}.
\end{align*}

For an adaptive algorithm, the selected set $E(H)$ is no longer fixed on every input. Hence, one could no longer express it as a clean Ising partition function. We instead analyze the KL divergence between the transcript distributions one query at a time.

\subsection{A cubic adaptive lower bound}

Fix a deterministic adaptive algorithm making $q$ distinct pair queries.
Let $T_t$ be its transcript after $t$ queries, and write $T=T_q$.
We will show that the transcript distributions $\mcald_{1,T}$ and
$\mcald_{0,T}$ under the yes and no distributions are statistically close when
$q\ll1/\rho^3$. Unless specified, expectations below are taken over the joint yes distribution of the hidden partition and the graph.

Our first step is to express the KL divergence between $\mcald_{1,T}$ and
$\mcald_{0,T}$ as the sum of expected squares of prediction advantages of each step.
At step $t$, the queried pair $E_t=(U_t,V_t)$ is determined by
$T_{t-1}$; let $Y_t$ be its answer. We view this step as a prediction
of the hidden product
$$
 X_t=\Sigma_{U_t}\Sigma_{V_t},\qquad
 C_t=\E[X_t\mid T_{t-1}].
$$
Given the history $T_{t-1}$, the next answer is a fair sign under the
no distribution. Under the yes distribution, its mean is $\rho C_t$:
conditioned also on $\Sigma$, its mean is $\rho X_t$, and averaging over
$\Sigma$ gives $\rho C_t$. Let $d(z)$ be the KL divergence of a
$\{\pm1\}$ variable with mean $z$ from a fair sign. The KL chain rule
therefore gives, for $\rho\le1/2$,
$$
 \begin{aligned}
 \KL(\mcald_{1,T}\Vert\mcald_{0,T})
 &=\sum_{t=1}^q\E[d(\rho C_t)]\\
 &\le\tfrac23(\log e)\rho^2\sum_{t=1}^q\E[C_t^2],
 \end{aligned}
$$
using $d(z)\le\tfrac23(\log e)z^2$ for $|z|\le1/2$.
Thus it suffices to bound the combined strength of the algorithm's
predictions before the corresponding answers are seen.

The difficulty is that each $C_t$ refers to a different history, and
the queried pairs themselves depend on those histories. We put all
predictions under the same conditional distribution by defining
$$
 R_t=\E[X_t\mid T].
$$
Since $T$ contains $T_{t-1}$, we have $C_t=\E[R_t\mid T_{t-1}]$. 
Jensen's inequality gives
$$
 \sum_{t=1}^q\E[C_t^2]
 =\sum_{t=1}^q\E\left[\bigl(\E[R_t\mid T_{t-1}]\bigr)^2\right]
 \le\sum_{t=1}^q\E[R_t^2].
$$

To see what this buys us, fix a complete transcript $h$.
Let $E(h)$ be
its set of queried pairs, and let $\mu_h$ be the conditional distribution
of $\Sigma$ given $T=h$. Define
$$
 S(h):=\sum_{t=1}^qR_t(h)^2
 =\sum_{(u,v)\in E(h)}
       \bigl(\E_{\mu_h}[\Sigma_u\Sigma_v]\bigr)^2.
$$
The selected pair set $E(h)$ is now fixed, and every term in $S(h)$
is evaluated under the same distribution $\mu_h$. Consequently,
\begin{equation}\label{eq:overview-adaptive-reduction}
 \KL(\mcald_{1,T}\Vert\mcald_{0,T})
 \le\tfrac23(\log e)\rho^2\E[S(T)].
\end{equation}
Thus, the remaining task is to show that $S(T)$ is small in expectation. We show it in two steps: we upper bound every $S(h)$ by a degree-two polynomial of $k(h):=\KL(\mu_h\|\Unif_n)$, where $\Unif_n$ is the uniform distribution on $\{\pm1\}^n$; then we bound its first two moments.

We prove in \Cref{lem:leveltwo} that
for every distribution $\mu_h$ and every pair set $E(h)$ that defines $S(h)=\sum_{(u,v)\in E(h)}\bigl(\E_{\Sigma\sim\mu_h}[\Sigma_u\Sigma_v]\bigr)^2$,
$$
 S(h)=O\bigl(k(h)+k(h)^2\bigr).
$$
To see the idea behind it, collect the pair correlations $r_{uv}=\E_{\mu_h}[\Sigma_u\Sigma_v]$ into the quadratic function $g(x)=\sum_{(u,v)\in E(h)}r_{uv}x_ux_v$. The choice of coefficients makes its mean under $\mu_h$ and its variance under $\Unif_n$ both equal to $S(h)$: $\E_{x\sim\mu_h}[g(x)]=\E_{x\sim \Unif_n}[g(x)^2]=S(h)$. The entropy variational inequality~\eqref{eq:duality} compares the former with an exponential moment under $\Unif_n$, and degree-two hypercontractivity controls that moment using the latter. This gives the stated bound in terms of $k(h)$.

It remains to bound the moments of $K=k(T)$ over the random
transcript $T$. We note that its first moment is exactly the mutual information between the
hidden partition and the transcript. By the fact that each query answer reveals at most $\rho^2$ information about the hidden partition,
$$
 \E[K]=I(\Sigma;T)
 =\sum_{t=1}^q I(\Sigma;Y_t\mid T_{t-1})
 =O(q\rho^2).
$$
To bound its second moment $\E[K^2]$, we use the information density
$$
 J=\log\frac{\mu_T(\Sigma)}{2^{-n}}
   =\sum_{t=1}^q J_t,
 \qquad
 J_t=\log\frac{1+\rho Y_tX_t}{1+\rho Y_tC_t},
$$
for which $K=\E[J\mid T]$. By Jensen's inequality, $\E[K^2]\le \E[J^2]$. Each $J_t$ is the log-likelihood ratio of the $t$-th query answer:
$$J_t=\log\frac{\Pr(Y_t\mid\Sigma,T_{t-1})}{\Pr(Y_t\mid T_{t-1})}.$$
Our strategy is to decompose each increment $J_t$ into a drift part $A_t:=\E[J_t\mid\Sigma,T_{t-1}]$ and a martingale part $J_t-A_t$. The drift part $A_t$ is exactly the KL-divergence between the distribution of the query answer $Y_t$ conditioned on $(\Sigma,T_{t-1})$ and its marginal distribution given $T_{t-1}$ alone. The martingale part has a zero mean even conditioned on $(\Sigma,T_{t-1})$. Both parts are bounded cleanly in~\Cref{lem:infodensity}. We obtain
$$\E[K]\le O(q\rho^2),\qquad\E[K^2]\le O(q\rho^2+q^2\rho^4).$$

Combining them with \eqref{eq:overview-adaptive-reduction} and
\Cref{lem:leveltwo}, we obtain
$$
 \begin{aligned}
 \KL(\mcald_{1,T}\Vert\mcald_{0,T})
 &\lesssim\rho^2\bigl(\E[K]+\E[K^2]\bigr)\\
 &=O\bigl(q\rho^4+q^2\rho^6\bigr).
 \end{aligned}
$$
Thus, when $q\le c/\rho^3$ for a sufficiently small constant $c$, this divergence is small. We prove that any algorithm that estimates MAX-CUT with high probability must pay query complexity $\Omega(1/\eps^3)$.

\section{Preliminaries}\label{sec:model}

We always denote $[k]$ as the set $\{1,\dots, k\}$.

\paragraph{Adjacency matrix (dense graph) model.}
We work in the \emph{adjacency matrix model}, also known as the
\emph{dense graph model} in property testing literature. The input is a simple
graph $G=(V,E)$. An algorithm accesses $G$ through \emph{pair queries}:
a query specifies an unordered pair $(u,v)\in\binom{V}{2}$
and returns $1$ if $(u,v)\in E$ and $0$ otherwise.
The \emph{query transcript} after $t$ queries is the ordered
sequence of the first $t$ queried pairs and their answers.

Algorithms may be randomized and have unrestricted computation.
Between queries, an algorithm may perform arbitrary computation
using its internal randomness and the current query transcript.
Its final output may depend on its internal randomness and the
complete transcript.
We consider both \emph{adaptive} algorithms, whose next query may
depend on the current query transcript, and \emph{non-adaptive} algorithms,
whose queries are determined by their internal randomness before
any answers are observed.

The \emph{worst-case query complexity} is the maximum number of
distinct pairs queried, over all input graphs and all outcomes
of the algorithm's internal randomness.

\paragraph{MAX-CUT.}

Given as input a simple graph $G=(V,E)$, we represent a cut of $V$ as a vector $\sigma\in \{-1,1\}^n$, where $\sigma_v$ denotes the part assignment to vertex $v$. The \emph{cut size} of $\sigma$ is the number of edges crossing between the two parts, i.e.,
$$\cutg(\sigma):=\left|\{(u,v)\in E:\sigma_u\ne \sigma_v\}\right|.$$
We also define the \emph{max cut size} of $G$ as
$$\MC(G):=\max_\sigma\cutg(\sigma).$$
For a fixed error parameter $\varepsilon\in(0,1]$, the \emph{MAX-CUT} problem asks for an approximate value $\hat M$ such that $|\hat M-\MC(G)|\le \varepsilon n^2$.
For this problem, a randomized tester must return such a value
with probability at least $2/3$ on every input graph.

\paragraph{MAX-2CSP.}
An input instance $F$ of MAX-2CSP consists of an integer $n$ and a collection of distinct triples $(f_i,u_i,v_i), f_i:\{0,1\}^2\rightarrow\{0,1\}, 1\le u_i<v_i\le n$.
A pair query $(u,v)$ on this instance reveals all constraints between $x_{u}$ and $x_{v}$.
The MAX-2CSP value $\mathrm{OPT}(F)$ is defined as the maximum number of constraints that are simultaneously satisfiable:
$$\mathrm{OPT}(F):=\max_{x\in\{0,1\}^n}\sum_{i}f_i(x_{u_i},x_{v_i}).$$
For a fixed error parameter $\varepsilon\in (0,1]$, the \emph{MAX-2CSP} problem asks for an approximate value $\hat M$ such that $|\hat M-\mathrm{OPT}(F)|\le \eps n^2$. A randomized tester must return such a value with probability at least $2/3$ on every input.

In this problem, we use $F[Q], Q\subseteq [n]$ to denote the set of constraints whose two variables both belong to $Q$. Since all constraints are distinct, there are at most $2^4=16$ constraints between any pair of variables.

\paragraph{Probability and information theory.}
Throughout, $\log$ denotes the base-$2$ logarithm. KL divergence,
entropy, and mutual information are therefore measured in bits.
When a base-$2$ logarithm is applied to an ordinary exponential,
$\log(e^x)=(\log e)x$.

We write $\Pr$ for probability; a subscript specifies the sampling
distribution when needed.
We write $X\overset{\mathrm d}{=}Y$ to mean that $X$ and $Y$
have the same distribution.
We use $\mcald$ for input distributions, their joint generative
distributions, and distributions of algorithm observations. Greek
letters denote distributions of the hidden partition and their comparison
distributions.
We use $\Unif_m$ to denote the uniform distribution on $\{-1,1\}^m$.

For a discrete distribution $\mcald$, write
$\mcald(x):=\Pr_{X\sim\mcald}[X=x]$.
For two distributions $\mcald,\mcald'$ on the same discrete
space $\Omega$, their \emph{total variation distance} is
$$
\TV(\mcald,\mcald')
:=\frac12\sum_{x\in\Omega}|\mcald(x)-\mcald'(x)|.
$$

Assuming that $\mcald'(x)>0$ whenever $\mcald(x)>0$, the \emph{likelihood ratio}
of $\mcald$ relative to $\mcald'$ is
$L(x):=\mcald(x)/\mcald'(x)$ for
$x\in\operatorname{supp}(\mcald')$.
Their \emph{$\chi^2$-divergence} and
\emph{Kullback--Leibler (KL) divergence} are defined by
$$
\chi^2(\mcald\|\mcald')
:=\E_{X\sim\mcald'}[L(X)^2]-1,
\qquad
\KL(\mcald\|\mcald')
:=\sum_{x\in\Omega}\mcald(x)
\log\frac{\mcald(x)}{\mcald'(x)}.
$$
Here, terms with $\mcald(x)=0$
contribute zero to the KL divergence.

We use the following standard inequalities.

\begin{proposition}[\cite{canonne2022topics}]
\label{prop:tvd_chi2}
For any two distributions $\mcald,\mcald'$ on the same
discrete space satisfying
$\mcald'(x)>0$ whenever $\mcald(x)>0$, we have
$$
\TV(\mcald,\mcald')
\le\frac12\sqrt{\chi^2(\mcald\|\mcald')},
\qquad
\TV(\mcald,\mcald')
\le\sqrt{\frac{\KL(\mcald\|\mcald')}{2\log e}}.
$$
\end{proposition}

We also use the elementary bound
\begin{equation}\label{eq:klchi2}
 \KL(\mcald\Vert\mcald')\le(\log e)\chi^2(\mcald\Vert\mcald'),
\end{equation}
for distributions on the same discrete space with
$\mcald'(x)>0$ whenever $\mcald(x)>0$; it follows from
$\log x\le(\log e)(x-1)$.
Conditional Jensen gives, for any convex $\varphi$ and random variables
$X,Y$, that $\varphi(\E[X\mid Y])\le\E[\varphi(X)\mid Y]$; we use the
case $\varphi(x)=x^2$, namely $(\E[X\mid Y])^2\le\E[X^2\mid Y]$.

Let $\mcald,\mcald'$ be distributions on the finite product space
$\Omega_1\times\cdots\times\Omega_q$, with $\mcald'(x)>0$
whenever $\mcald(x)>0$. Their KL-divergence satisfies a chain rule:
$$
\KL(\mcald\|\mcald')
=
\sum_{t=1}^q
\E_{X\sim\mcald}
\left[
\KL\bigl(
\mcald_t(\cdot\mid X_{<t})
\|
\mcald'_t(\cdot\mid X_{<t})
\bigr)
\right],
$$
where $X_{<t}=(X_1,\ldots,X_{t-1})$, and $\mcald_t$
and $\mcald'_t$ denote the conditional distributions of
coordinate $t$ given the prefix under $\mcald$ and $\mcald'$,
respectively.

\begin{proposition}[Entropy variational inequality]
\label{prop:entropy_variational}
Let $\mcald,\mcald'$ be distributions on a finite set $\Omega$, with
$\mcald'(x)>0$ whenever $\mcald(x)>0$. For every function $f:\Omega\to\R$,
\begin{equation}\label{eq:duality}
 \E_{\mcald} [f]\le\frac{\KL(\mcald\Vert\mcald')
 +\log\E_{\mcald'} [e^f]}{\log e}.
\end{equation}
\end{proposition}
\begin{proof}
Set $z=\E_{\mcald'} [e^f]$ and define $\mcald'_f(x)=\mcald'(x)e^{f(x)}/z$.
Nonnegativity of KL divergence gives
$$
0\le\KL(\mcald\Vert\mcald'_f)
=\KL(\mcald\Vert\mcald')-(\log e)\E_{\mcald} [f]+\log z.
$$
Rearranging proves the inequality.
\end{proof}


For random variables $X,Y$, let $\mcald_X$ and
$\mcald_{XY}$ denote their marginal and joint distributions.
For each $y$ with $\mcald_Y(y)>0$, let $\mcald_{X\mid Y=y}$
denote the conditional distribution of $X$ given $Y=y$.
The \emph{entropy} of $X$ and the \emph{conditional entropy}
of $X$ given $Y$ are
$$
\begin{aligned}
H(X)
&:=-\sum_{x:\mcald_X(x)>0}
\mcald_X(x)\log\mcald_X(x),\\
H(X\mid Y)
&:=-\sum_{x,y:\mcald_{XY}(x,y)>0}
\mcald_{XY}(x,y)\log\mcald_{X\mid Y=y}(x).
\end{aligned}
$$
Their \emph{mutual information} is
$$
I(X;Y)
:=\KL(\mcald_{XY}\|\mcald_X\otimes\mcald_Y)
=H(X)-H(X\mid Y),
$$
where $\mcald_X\otimes\mcald_Y$ is the product of the two
marginal distributions. Equivalently,
$$
I(X;Y)
=
\E_{y\sim\mcald_Y}
\left[
\KL(\mcald_{X\mid Y=y}\|\mcald_X)
\right].
$$
Mutual information measures the dependence between $X$ and $Y$.

The \emph{conditional mutual information} of $X$ and $Y$
given $Z$ is
$$
I(X;Y\mid Z):=H(X\mid Z)-H(X\mid Y,Z).
$$
Mutual information satisfies the chain rule
$$
I(X;Y_1,\ldots,Y_q)
=
\sum_{t=1}^q I(X;Y_t\mid Y_1,\ldots,Y_{t-1}).
$$

For scalar random variables $X,Y$ with finite second moments,
$\Cov(X,Y):=\E[(X-\E [X])(Y-\E [Y])]$ denotes their covariance,
and $\Var(X):=\E[(X-\E [X])^2]$ denotes the variance of $X$.
Expectations are under the specified joint distribution; a subscript
indicates that distribution when needed.

\paragraph{Matrices, norms, and derivatives.}
For a vector $v\in\R^n$, let
$\|v\|_2=(\sum_i v_i^2)^{1/2}$ denote its Euclidean norm.
For a matrix $B\in\R^{m\times n}$, its operator norm and
Frobenius norm are
$$
\|B\|_{\mathrm{op}}
:=\sup_{\|v\|_2=1}\|Bv\|_2,
\qquad
\|B\|_F
:=\left(\sum_{i=1}^m\sum_{j=1}^n B_{ij}^2\right)^{1/2}.
$$
Unless otherwise specified, $\|\cdot\|$ denotes the Euclidean
norm for vectors and the operator norm for matrices.
For a vector $v$, $\diag(v)$ is the diagonal matrix with
diagonal $v$. For a square matrix $B$, $\diag(B)$ retains
its diagonal entries and sets all other entries to zero.
For a twice differentiable scalar function $f:U\to\R$ on an open set
$U\subseteq\R^m$, its
\emph{gradient} is $\nabla f=(\partial f/\partial x_i)_i$ and its
\emph{Hessian} is $\nabla^2 f=(\partial^2 f/\partial x_i\partial x_j)_{i,j}$.
Subscripts indicate the variables being differentiated.

\paragraph{Concentration and moment bounds.}
We collect the tail, variance, and moment bounds used below.
The bounds for independent variables also apply after conditioning on
information that leaves the specified variables independent and
preserves the stated hypotheses.

\begin{proposition}[Hoeffding's inequality {\citep[Section 2.6]{BoucheronLugosiMassart2013}}]
\label{prop:hoeffding}
Let $X_1,\ldots,X_m$ be independent real random variables with
$a_i\le X_i\le b_i$ almost surely. Put $S=\sum_iX_i$ and
$v=\sum_i(b_i-a_i)^2$. If $v>0$, then, for every $t>0$,
$$
 \Pr(S-\E [S]\ge t)\le e^{-2t^2/v},\qquad
 \Pr(S-\E [S]\le-t)\le e^{-2t^2/v}.
$$
Consequently $\Pr(|S-\E [S]|\ge t)\le2e^{-2t^2/v}$.
The exponential-moment bound behind these inequalities is
$$
 \E [e^{\lambda(X_i-\E [X_i])}]
 \le e^{\lambda^2(b_i-a_i)^2/8}\qquad(\lambda\in\R).
$$
\end{proposition}

The following multiplicative bound is a standard consequence of the
Cram\'er--Chernoff method \citep[Section 2.2]{BoucheronLugosiMassart2013}.
\begin{proposition}[Multiplicative Chernoff bound]\label{prop:chernoff}
Let $X_1,\ldots,X_m$ be independent random variables in $[0,1]$,
and put $S=\sum_iX_i$ and $\mu=\E [S]>0$. For every $\delta>0$,
$$
 \Pr(S\ge(1+\delta)\mu)
 \le\exp\left(-\frac{\delta^2\mu}{2+\delta}\right).
$$
\end{proposition}

The next inequality bounds variance through the effect of independently
resampling each input coordinate.
\begin{proposition}[Efron--Stein inequality {\citep[Section 3.1]{BoucheronLugosiMassart2013}}]
\label{prop:efron-stein}
Let $X=(X_1,\ldots,X_m)$ have independent coordinates in spaces
$\Omega_1,\ldots,\Omega_m$, and let $f:\prod_i\Omega_i\to\R$ satisfy
$\E[f(X)^2]<\infty$. For each $i$, let $X_i'$ have the same distribution
as $X_i$ and be independent of $X$, and set
$X^{(i)}=(X_1,\ldots,X_{i-1},X_i',X_{i+1},\ldots,X_m)$. Then
$$
 \Var(f(X))\le\frac12\sum_{i=1}^m
 \E\bigl[(f(X)-f(X^{(i)}))^2\bigr].
$$
\end{proposition}

For adaptive observations, we use the following second-moment identity.
A filtration $(\mathcal F_t)_{t=0}^q$ is an increasing sequence of
sigma-algebras describing the information available at each step.
\begin{proposition}[Orthogonality of martingale increments]
\label{prop:martingale-orthogonality}
Let $(\mathcal F_t)_{t=0}^q$ be a filtration, and let $D_1,\ldots,D_q$
be square-integrable real random variables such that $D_t$ is
$\mathcal F_t$-measurable and $\E[D_t\mid\mathcal F_{t-1}]=0$.
Then $\E[D_sD_t]=0$ for $1\le s<t\le q$, and
$$
 \E\left[\left(\sum_{t=1}^q D_t\right)^2\right]
 =\sum_{t=1}^q\E[D_t^2].
$$
\end{proposition}
\begin{proof}
For $s<t$, the variable $D_s$ is $\mathcal F_{t-1}$-measurable, so
$$
 \E[D_sD_t]
 =\E\left[D_s\E[D_t\mid\mathcal F_{t-1}]\right]=0.
$$
Expanding the square and using these zero cross terms gives the identity.
\end{proof}

For a nonnegative random variable $Z$ and $a>0$, Markov's inequality
is $\Pr(Z\ge a)\le\E [Z]/a$.

\section{A general quartic non-adaptive upper bound}\label{sec:upper}

In this section, we give an $O(1/\varepsilon^4)$ non-adaptive upper bound for MAX-CUT. This is achieved by a canonical tester. Our proof combines the cut-norm estimate for random submatrices of \citet{RudelsonVershynin2007} with a geometric comparison of weighted feasible moment sets.

\begin{theorem}[Quartic query upper bound]\label{thm:upper}
For every $n\ge2$ and $0<\eps\le1$, there is a randomized non-adaptive
tester for MAX-CUT on $n$-vertex graphs with error parameter $\eps$ and
success probability at least $2/3$ that makes $O(1/\eps^4)$ pair
queries in the worst case. Its running time is $2^{O(1/\eps^2)}$ in the
RAM model.
\end{theorem}

We derive this bound from the canonical tester below: sample a subset
of vertices, query the induced subgraph, and rescale its MAX-CUT value.
By \citet[Lemma~4.1]{GoldreichTrevisan2003}, any adaptive tester for a
dense graph property can be simulated by such a non-adaptive tester
with a quadratic query blowup, so a canonical tester loses at most a
quadratic factor against the best adaptive algorithm. The main work of this section is to prove that the canonical tester
already succeeds with $s=O(1/\eps^2)$ sampled vertices, hence
$O(1/\eps^4)$ pair queries. The matching lower bound in
\Cref{thm:na} makes this optimal among non-adaptive algorithms.
\begin{center}
\begin{minipage}{0.94\linewidth}
\raggedright
\hrule\smallskip
\refstepcounter{algo}\label{alg:canonical}%
\textbf{Algorithm \thealgo} Canonical tester for \Cref{thm:upper}.
\begin{itemize}[leftmargin=1.5em,itemsep=2pt,topsep=4pt,parsep=0pt,partopsep=0pt]
 \item Set $s=\lceil C/\eps^2\rceil$ for a sufficiently large absolute constant $C$.
 \item If $n\le s$, query all $\binom n2$ pairs and return $\MC(G)$.
 \item Otherwise include each vertex in $Q$ independently with probability $s/n$.
 \begin{itemize}[leftmargin=1.2em,itemsep=2pt,topsep=2pt,parsep=0pt,partopsep=0pt]
  \item If $|Q|>2s$, return zero without making queries.
  \item Otherwise query all pairs in $Q$ and return $n^2\MC(G[Q])/s^2$.
 \end{itemize}
\end{itemize}
\smallskip\hrule
\end{minipage}
\end{center}

What remains is to show that the above canonical tester outputs a good estimate of the MAX-CUT value of the full graph with high probability.

\begin{theorem}\label{thm:sampling}
There is an absolute constant $C_0>0$ such that the following holds.
Let $1\le s\le n$ be real, and sample a subset $Q$ by independently including each vertex with probability $s/n$. Then
$$
 \E\left[\left|\frac{\MC(G[Q])}{s^2}-\frac{\MC(G)}{n^2}\right|\right]
 \le\frac{C_0}{\sqrt s}.
$$
\end{theorem}

Most of our efforts in this section are devoted to proving \Cref{thm:sampling}. The upper bound \Cref{thm:upper} follows directly from it.

For the adjacency matrix $A$ of $G$ and the all-ones vector $\one$,
every cut $\sigma\in\{-1,1\}^n$ satisfies
$\cutg(\sigma)=\frac14(\one^TA\one-\sigma^TA\sigma)$, so
\begin{equation}\label{eq:cutidentity}
 \MC(G)=\max_{\sigma}\cutg(\sigma)
 =\frac14\Bigl(\one^TA\one-\min_{\sigma\in\{-1,1\}^n}\sigma^TA\sigma\Bigr).
\end{equation}

Our strategy is to apply a spectral truncation to $A$, and then to
compare the resulting quadratic minima by discarding $R$ on both
sides. Let $\theta_j$ and $u_j$ be the eigenvalues and orthonormal
eigenvectors of $A$, fix a threshold $0<\tau\le1$ to be determined
later, and write
$$
 L=\sum_{|\theta_j|>\tau n}\theta_j u_ju_j^T,\qquad R=A-L.
$$
Thus $L$ retains the eigenvalues of large absolute value, while
$\norm R_{\mathrm{op}}\le\tau n$ and $\norm R_F\le\norm A_F\le n$.

For the sampling analysis, fix $1\le s\le n$, set $p=s/n$, and let
$\xi_i\sim\mathrm{Bernoulli}(p)$ independently. Write
$P=\diag(\xi_1,\ldots,\xi_n)$ and $Q=\{i:\xi_i=1\}$. The sampled
adjacency matrix $A_{QQ}$ is indexed by $Q$; the $n\times n$ matrix
$PAP$ agrees with $A$ on $Q\times Q$ and is zero elsewhere.
Unless stated otherwise, expectations in this section are over the
vertex selectors.

We work with fractional vectors $x\in[-1,1]^n$. Since $A$ and $PAP$
have zero diagonal, their quadratic minima over the cube equal their
minima over $\{-1,1\}^n$: with all other coordinates fixed, the
objective is linear in each coordinate, so each coordinate can be
moved to an endpoint without increasing the value. All minima below
are over $x\in[-1,1]^n$.

The cut identity \eqref{eq:cutidentity} splits the expected sampling
error into an edge-density error and a quadratic-minimum error:
\begin{align}
 \E\left[\left|\frac{\MC(G[Q])}{s^2}-\frac{\MC(G)}{n^2}\right|\right]
 &\le\frac14\E\left[\left|
 \frac{\one^TPAP\one}{s^2}-\frac{\one^TA\one}{n^2}
 \right|\right]\nonumber\\
 &\quad+\frac14\E\left[\left|
 \frac{\min_x x^TPAPx}{s^2}-\frac{\min_x x^TAx}{n^2}
 \right|\right].
 \label{eq:upper-sampling-error}
\end{align}
We compare the quadratic minima by discarding $R$ on both sides and
comparing the two problems involving $L$:
\begin{align}
 \left|\frac{\min_x x^TPAPx}{s^2}-\frac{\min_x x^TAx}{n^2}\right|
 &\le\left|
 \frac{\min_x x^TPAPx}{s^2}-\frac{\min_x x^TPLPx}{s^2}
 \right|\nonumber\\
 &\quad+\left|
 \frac{\min_x x^TPLPx}{s^2}-\frac{\min_x x^TLx}{n^2}
 \right|\nonumber\\
 &\quad+\left|
 \frac{\min_x x^TLx}{n^2}-\frac{\min_x x^TAx}{n^2}
 \right|.
 \label{eq:upper-error-decomposition}
\end{align}
The first term is the cost of discarding $R$ in the sample. It is at
most $\frac{1}{s^2}\sup_{x\in[-1,1]^n}|x^TPRPx|$, and the cut-norm estimate
of \citet{RudelsonVershynin2007} bounds this largest sampled
contribution over all cuts at once, without a union bound over
individual cuts; \Cref{subsec:residual} carries this out. The second
term compares the sampled and full-graph minima for $L$. We relax it
geometrically: the full graph and the sample define two compact convex
sets, and \Cref{subsec:weighted} bounds their Hausdorff distance,
which bounds the gap between their minima without a factor depending on
the number of retained eigenvectors. The last term is at most $\tau$, by
$\norm R_{\mathrm{op}}\le\tau n$ and $\norm{x}_2^2\le n$. In
\Cref{subsec:upper-assembly}, we bound the first term on the right-hand
side of \eqref{eq:upper-sampling-error} by $3/(4\sqrt s)$ and combine
these estimates. Taking $\tau=1/\sqrt s$ then
proves \Cref{thm:sampling}.

\subsection{The cost of discarding $R$ in the sample}\label{subsec:residual}
In this subsection we bound the first term on the right-hand side of \eqref{eq:upper-error-decomposition}.

\begin{lemma}[Sampling the residual]\label{lem:residual-sampling}
There is an absolute constant $C_{\mathrm{res}}>0$ such that
$$
 \frac1{s^2}\E\left[\sup_{x\in[-1,1]^n}|x^TPRPx|\right]
 \le C_{\mathrm{res}}\left(\tau+\frac1{\tau s}+\frac1{\sqrt s}\right).
$$
\end{lemma}

Using only the facts that
$\norm{R_{QQ}}_{\mathrm{op}}\le\tau n$ and $\E[|Q|]=s$ gives
a loose bound of $\tau n/s$, which
depends on $n$.

To obtain a bound independent of $n$, we use the cut-norm estimate
for random submatrices of
\citet[Theorem~1.5]{RudelsonVershynin2007}.
We first define the two matrix norms used in this estimate. For a real
$n\times n$ matrix $B$, its $\ell_\infty\to\ell_1$ operator norm is
the largest possible value of $\sum_i|(Bx)_i|$ when $x$ ranges over
$[-1,1]^n$:
$$
 \norm{B}_{\infty\to1}
 :=\sup_{x\in[-1,1]^n}\sum_i|(Bx)_i|
 =\sup_{x,y\in[-1,1]^n}|y^TBx|.
$$
For every $x\in[-1,1]^n$, the definition gives
$|x^TBx|\le\norm{B}_{\infty\to1}$.
Thus this norm bounds $|\sigma^TB\sigma|$ simultaneously for every
cut $\sigma\in\{-1,1\}^n$.

The column norm $\|\cdot\|_{\mathrm{Col}}$ sums the
Euclidean lengths of the columns:
$$
 \norm{B}_{\mathrm{Col}}
 :=\sum_j\norm{B_{\cdot j}}_2
 =\sum_j\left(\sum_iB_{ij}^2\right)^{1/2}.
$$
Thus $\norm{B^T}_{\mathrm{Col}}$ sums the row lengths.
Cauchy--Schwarz gives $\norm{B}_{\mathrm{Col}}\le\sqrt n\,\norm{B}_F$.

Rudelson and Vershynin state their theorem for the cut norm, which is
equivalent to $\norm{\cdot}_{\infty\to1}$ up to a factor of four.\footnote{The
cut norm is $\norm{B}_{\square}:=\max_{I,J\subseteq[n]}
|\sum_{i\in I,j\in J}B_{ij}|$, the largest absolute sum over a rectangle
of entries. It satisfies $\frac14\norm{B}_{\infty\to1}\le
\norm{B}_{\square}\le\norm{B}_{\infty\to1}$.}
We restate their estimate using the
$\ell_\infty\to\ell_1$ operator norm, as in their proof.
\begin{lemma}[Rudelson--Vershynin cut-norm estimate {\citep[Theorem~1.5]{RudelsonVershynin2007}}]\label{lem:restriction}
Let $B$ be a deterministic real $n\times n$ matrix and let
$P=\diag(\xi_1,\ldots,\xi_n)$, where the $\xi_i$ are independent
Bernoulli$(p)$ variables with $0\le p\le1$. Then
$$
 \E[\norm{PBP}_{\infty\to1}]\le C_{\mathrm{RV}}\left(
 p^2\norm{B-\diag(B)}_{\infty\to1}+p\sum_i|B_{ii}|
 +p^{3/2}\bigl(\norm{B}_{\mathrm{Col}}+\norm{B^T}_{\mathrm{Col}}\bigr)\right),
$$
where $C_{\mathrm{RV}}>0$ is an absolute constant.
\end{lemma}

\begin{proof}[Proof of \Cref{lem:residual-sampling}]
For every realization of the selectors and every $x\in[-1,1]^n$,
$$
 |x^TPRPx|
 =\left|\sum_i x_i(PRPx)_i\right|
 \le\sum_i|(PRPx)_i|
 \le\norm{PRP}_{\infty\to1}.
$$
The inequalities use $|x_i|\le1$ and the definition of
$\norm{\cdot}_{\infty\to1}$.
Taking the supremum over $x$ and then expectations gives
\begin{equation}\label{eq:residual-reduction}
 \frac1{s^2}\E\left[\sup_{x\in[-1,1]^n}|x^TPRPx|\right]
 \le\frac{\E[\norm{PRP}_{\infty\to1}]}{s^2}.
\end{equation}
It therefore suffices to bound the right-hand side.
Set $H=R-\diag(R)$. Applying \Cref{lem:restriction} to $R$ and
using $p=s/n$ gives
\begin{equation}\label{eq:rv-normalized}
 \frac{\E[\norm{PRP}_{\infty\to1}]}{s^2}
 \le C_{\mathrm{RV}}\left(
 \frac{\norm{H}_{\infty\to1}}{n^2}
 +\frac{\sum_i|R_{ii}|}{ns}
 +\frac{\norm{R}_{\mathrm{Col}}+\norm{R^T}_{\mathrm{Col}}}
 {n^{3/2}\sqrt s}\right).
\end{equation}
We now bound the three terms in this estimate.

We first control the diagonal. Although $A$ has zero diagonal,
$L$ and $R$ need not. Since $R_{ii}=-L_{ii}$,
\begin{equation}\label{eq:residual-diagonal}
 |R_{ii}|=|L_{ii}|\le\sum_{j\text{ retained}}|\theta_j|u_j(i)^2
 \le\frac1{\tau n}\sum_{j\text{ retained}}\theta_j^2u_j(i)^2
 \le\frac{(A^2)_{ii}}{\tau n}\le\frac1\tau.
\end{equation}
To bound $\norm{H}_{\infty\to1}$, note that every $x\in[-1,1]^n$ has
$\norm{x}_2\le\sqrt n$, so Cauchy--Schwarz gives
$$
 \norm{Rx}_1\le\sqrt n\,\norm{Rx}_2
 \le\sqrt n\,\norm{R}_{\mathrm{op}}\norm{x}_2
 \le n\,\norm{R}_{\mathrm{op}}.
$$
Applying the triangle inequality to $Hx=Rx-\diag(R)x$ and taking
the supremum over $x$, we obtain
\begin{align}
 \norm{H}_{\infty\to1}
 &\le n\,\norm{R}_{\mathrm{op}}+\sum_i|R_{ii}|
 \le\tau n^2+n/\tau,\notag\\
 \norm{R}_{\mathrm{Col}},\,\norm{R^T}_{\mathrm{Col}}
 &\le\sqrt n\,\norm{R}_F\le n^{3/2}.
\label{eq:residual-norm-estimates}
\end{align}
Substituting \eqref{eq:residual-diagonal} and \eqref{eq:residual-norm-estimates}
into \eqref{eq:rv-normalized} gives
\begin{align}
 \frac{\E[\norm{PRP}_{\infty\to1}]}{s^2}
 &\le C_{\mathrm{RV}}\left(
 \tau+\frac1{\tau n}+\frac1{\tau s}+\frac2{\sqrt s}\right)
 \nonumber\\
 &\le2C_{\mathrm{RV}}\left(
 \tau+\frac1{\tau s}+\frac1{\sqrt s}\right).
 \label{eq:residual}
\end{align}
The last inequality uses $s\le n$.
Together with \eqref{eq:residual-reduction}, \eqref{eq:residual} proves the lemma with
$C_{\mathrm{res}}=2C_{\mathrm{RV}}$.
\end{proof}

\subsection{The sampling error for $L$}\label{subsec:weighted}
In this subsection, we bound the second term on the right-hand side of \eqref{eq:upper-error-decomposition}.
Throughout this subsection, we write $\norm{v}=\norm{v}_2$ for the
Euclidean norm of a vector $v$.

\begin{lemma}\label{lem:retained-sampling}
\begin{equation}\label{eq:mincompare}
 \E\left[\left|
 \min_{x\in[-1,1]^n}\frac{x^TPLPx}{s^2}
 -\min_{x\in[-1,1]^n}\frac{x^TLx}{n^2}
 \right|\right]
 \le\frac8{\sqrt s}+\frac{17}{\tau s}.
\end{equation}
\end{lemma}

We will use the following standard contraction inequality.
\begin{lemma}[Rademacher contraction principle {\citep[Chapter~4]{LedouxTalagrand1991}}]
\label{lem:rademacher-contraction}
Let $n\ge1$, let $\mathcal A\subseteq\R^n$ be nonempty and bounded,
and let $\Xi_1,\ldots,\Xi_n$ be independent signs, each uniform on
$\{-1,1\}$. Then
$$
 \E_{\Xi}\left[\sup_{a\in\mathcal A}\sum_{i=1}^n\Xi_i|a_i|\right]
 \le\E_{\Xi}\left[\sup_{a\in\mathcal A}\sum_{i=1}^n\Xi_ia_i\right].
$$
\end{lemma}

To prove \Cref{lem:retained-sampling}, we first express the two
quadratic minima as minima of the same function over two convex sets.
For every $x\in[-1,1]^n$, the spectral decomposition of $L$ gives
\begin{align*}
 \frac{x^TLx}{n^2}
 &=\sum_{j\text{ retained}}\frac{\theta_j}{n^2}
   \left(\sum_i x_iu_j(i)\right)^2\\
 &=\sum_{j\text{ retained}}\frac{\theta_j}{n}
   \left(\frac1n\sum_i x_i\sqrt n\,u_j(i)\right)^2.
\end{align*}
The last expression writes the objective in terms of averages over
vertices. Accordingly, set $\lambda_j=\theta_j/n$ and
$\phi_j(i)=\sqrt n\,u_j(i)$, so that
\begin{equation}\label{eq:retained-coordinates}
 \frac{x^TLx}{n^2}
 =\sum_j\lambda_j\left(\frac1n\sum_i x_i\phi_j(i)\right)^2.
\end{equation}
Here and below, $j$ ranges over the retained eigenvalues. Since the
$u_j$ are orthonormal eigenvectors, the rescaled $\phi_j$ satisfy
$\frac1n\sum_i\phi_j(i)^2=1$ and $\frac1n\sum_i\phi_j(i)\phi_k(i)=0$
for $j\neq k$. Thus they are orthonormal with respect to the uniform
distribution on the vertices; also $|\lambda_j|>\tau$ for every $j$.

For each vertex $i$, define
the vector $y_i=(\lambda_j\phi_j(i))_j$, and set
$D=\diag(1/\lambda_j)$ and $f(w)=w^TDw=\sum_j w_j^2/\lambda_j$.
Equation~\eqref{eq:retained-coordinates} gives the exact identities
\begin{equation}\label{eq:moment-objectives}
 f\left(\frac1n\sum_i x_iy_i\right)=\frac{x^TLx}{n^2},\qquad
 f\left(\frac1s\sum_i \xi_ix_iy_i\right)=\frac{x^TPLPx}{s^2}.
\end{equation}
Therefore the two minima to be compared are the minima of $f$ over the
compact convex sets
$$
 K=\left\{\frac1n\sum_i x_iy_i:x\in[-1,1]^n\right\},\qquad
 K_Q=\left\{\frac1s\sum_i \xi_ix_iy_i:x\in[-1,1]^n\right\},
$$
and it remains to bound $|\min_{K_Q}f-\min_K f|$. We assume that at least one eigenvalue is retained as otherwise the claim is trivial.

We will compare $K_Q$ and $K$ using the Hausdorff distance.
For nonempty compact sets $C_1,C_2$ in a Euclidean space, their
\emph{Hausdorff distance} is
$$
 d_H(C_1,C_2):=\max\left\{
 \sup_{u\in C_1}\inf_{w\in C_2}\norm{u-w}_2,\;
 \sup_{w\in C_2}\inf_{u\in C_1}\norm{u-w}_2
 \right\}.
$$
Thus $d_H(C_1,C_2)\le\delta$ means that every point of either set
has a point of the other set within Euclidean distance $\delta$.
The \emph{support function} of a nonempty compact set $C$ is
$$
 h_C(v):=\sup_{w\in C}\ip{v}{w}.
$$
For a unit vector $v$, it is the largest projection of a point of $C$
onto $v$. If $C_1$ and $C_2$ are also convex, then
\begin{equation}\label{eq:hausdorff-support}
 d_H(C_1,C_2)=\sup_{\norm{v}_2\le1}|h_{C_1}(v)-h_{C_2}(v)|.
\end{equation}
Indeed, for $\delta\ge0$, the condition that every point of $C_1$
lies within distance $\delta$ of $C_2$ is equivalent, by the separating
hyperplane theorem, to $h_{C_1}(v)\le h_{C_2}(v)+\delta$ for all unit
vectors $v$. Applying this in both directions proves the identity.

The next lemma bounds the first two moments of this distance in terms
of the average squared length of the vectors $y_i$.
\begin{lemma}[Moments of the Hausdorff distance]\label{lem:hausdorff-moments}
For the sets $K,K_Q$ defined above, set
$$
 \Delta:=d_H(K_Q,K),\qquad t_y:=\frac1n\sum_i\norm{y_i}^2.
$$
Then
$$
 \E[\Delta]\le4\sqrt{t_y/s},\qquad
 \E[\Delta^2]\le\frac{17t_y}{s}.
$$
\end{lemma}
\begin{proof}
In our setting, the support function of $K$ is
$$
 h_K(v)=\sup_{x\in[-1,1]^n}\left\langle v,\frac1n\sum_i x_iy_i\right\rangle
 =\frac1n\sum_i|\ip v{y_i}|,
$$
because each coordinate $x_i$ is optimally set to the sign of
$\ip v{y_i}$; likewise $h_{K_Q}(v)=\frac1s\sum_i\xi_i|\ip v{y_i}|$.
Therefore \eqref{eq:hausdorff-support} gives
\begin{equation}\label{eq:hausdorff-selector-form}
 \Delta=\sup_{\norm v\le1}\left|\frac1n\sum_i|\ip v{y_i}|
 -\frac1s\sum_i\xi_i|\ip v{y_i}|\right|.
\end{equation}

We use a \emph{ghost-sample symmetrization} argument to bound
$\E[\Delta]$. To that end, we introduce independent
$\mathrm{Bernoulli}(p)$ selectors $\xi_1',\ldots,\xi_n'$,
independent of all the original selectors $\xi_i$. They form an
independent copy of the original sample, called the \emph{ghost sample}.
We also introduce independent Rademacher signs $\Xi_i\in\{-1,1\}$,
independent of both sets of selectors.

Since $p=s/n$ and $\E_{\xi'}[\xi_i']=p$, for each fixed realization
of the original selectors we have
\begin{align}
 \Delta
 &=\frac1s\sup_{\norm v\le1}
 \left|\sum_i(\xi_i-p)|\ip v{y_i}|\right|\notag\\
 &=\frac1s\sup_{\norm v\le1}
 \left|\E_{\xi'}\left[\sum_i(\xi_i-\xi_i')|\ip v{y_i}|\right]\right|\notag\\
 &\le\frac1s\E_{\xi'}\left[\sup_{\norm v\le1}
 \left|\sum_i(\xi_i-\xi_i')|\ip v{y_i}|\right|\right].
\label{eq:hausdorff-ghost}
\end{align}
The inequality follows by moving the absolute value and supremum
inside the expectation over the independent copy.

Swapping $\xi_i$ and
$\xi_i'$ preserves their joint distribution and reverses the sign
of $\xi_i-\xi_i'$. Since the pairs are independent across $i$,
we may flip each difference independently without changing their
joint distribution:
$$
 (\xi_i-\xi_i')_{i=1}^n
 \overset{\mathrm d}{=}
 (\Xi_i(\xi_i-\xi_i'))_{i=1}^n.
$$
Taking expectations in \eqref{eq:hausdorff-ghost} and then applying the
triangle inequality gives
\begin{align}
 \E[\Delta]
 &\le\frac1s\E_{\xi,\xi',\Xi}\left[\sup_{\norm v\le1}
 \left|\sum_i\Xi_i(\xi_i-\xi_i')|\ip v{y_i}|\right|\right]\notag\\
 &\le\frac1s\E_{\xi,\Xi}\left[\sup_{\norm v\le1}
 \left|\sum_i\Xi_i\xi_i|\ip v{y_i}|\right|\right]\notag\\
 &\quad+\frac1s\E_{\xi',\Xi}\left[\sup_{\norm v\le1}
 \left|\sum_i\Xi_i\xi_i'|\ip v{y_i}|\right|\right]\notag\\
 &=\frac2s\E_{\xi,\Xi}\left[\sup_{\norm v\le1}
 \left|\sum_i\Xi_i\xi_i|\ip v{y_i}|\right|\right],
\label{eq:hausdorff-symmetrization}
\end{align}
where the last equality uses that the two sets of selectors have
the same distribution.

To bound the last expression, fix the selectors and apply
\Cref{lem:rademacher-contraction} with
$\mathcal A=\{(\xi_i\ip v{y_i})_{i=1}^n:\norm v\le1\}$.
This set is bounded because $|\xi_i\ip v{y_i}|\le\norm{y_i}$.
Since $v=0$ is allowed, the suprema of the expression and its negative
are both nonnegative. We can therefore bound the supremum of the
absolute value by their sum:
\begin{align}
 &\E_{\Xi}\left[\sup_{\norm v\le1}
 \left|\sum_i\Xi_i\xi_i|\ip v{y_i}|\right|\right]\notag\\
 &\quad\le\E_{\Xi}\left[\sup_{\norm v\le1}
 \sum_i\Xi_i\xi_i|\ip v{y_i}|\right]
 +\E_{\Xi}\left[\sup_{\norm v\le1}
 \sum_i(-\Xi_i)\xi_i|\ip v{y_i}|\right]\notag\\
 &\quad=2\E_{\Xi}\left[\sup_{\norm v\le1}
 \sum_i\Xi_i\xi_i|\ip v{y_i}|\right]\notag\\
 &\quad\le2\E_{\Xi}\left[\sup_{\norm v\le1}
 \sum_i\Xi_i\xi_i\ip v{y_i}\right]\notag\\
 &\quad=2\E_{\Xi}\left[\sup_{\norm v\le1}
 \left\langle v,\sum_i\Xi_i\xi_iy_i\right\rangle\right]\notag\\
 &\quad=2\E_{\Xi}\left[\left\|\sum_i\Xi_i\xi_iy_i\right\|\right].
\label{eq:hausdorff-contraction}
\end{align}
The first equality uses $(\Xi_i)_i\overset{\mathrm d}{=}(-\Xi_i)_i$,
and the following inequality follows from
\Cref{lem:rademacher-contraction}.
The last equality uses $\sup_{\norm v\le1}\ip vz=\norm z$.
Averaging \eqref{eq:hausdorff-contraction} over the selectors and
combining it with \eqref{eq:hausdorff-symmetrization} gives
\begin{align}
 \E[\Delta]
 &\le\frac4s\,\E_{\xi,\Xi}\left[\left\|\sum_i\Xi_i\xi_iy_i\right\|\right]\notag\\
 &\le\frac4s\sqrt{\E_{\xi,\Xi}\left[\left\|\sum_i\Xi_i\xi_iy_i\right\|^2\right]}.
\label{eq:hausdorff-rademacher-bound}
\end{align}
The second inequality is Jensen's inequality for the concave square-root
function. To evaluate the expectation under the square root,
first condition on the selectors. The signs $\Xi_i$ remain independent
and uniform on $\{-1,1\}$. Expanding the square and using $\Xi_i^2=1$
gives
\begin{align*}
 \E_\Xi\left[\left\|\sum_i\Xi_i\xi_iy_i\right\|^2\right]
 &=\sum_i\xi_i^2\norm{y_i}^2
   +2\sum_{i<j}\xi_i\xi_j\ip{y_i}{y_j}\E_\Xi[\Xi_i\Xi_j]\\
 &=\sum_i\xi_i^2\norm{y_i}^2
   +2\sum_{i<j}\xi_i\xi_j\ip{y_i}{y_j}
     \underbrace{\E_\Xi[\Xi_i]\E_\Xi[\Xi_j]}_{=0}\\
 &=\sum_i\xi_i\norm{y_i}^2,
\end{align*}
where the last equality also uses $\xi_i^2=\xi_i$.
Taking expectation over the selectors then gives
\begin{align}
 \E_{\xi,\Xi}\left[\left\|\sum_i\Xi_i\xi_iy_i\right\|^2\right]
 &=\E_\xi\left[\sum_i\xi_i\norm{y_i}^2\right]\notag\\
 &=\sum_i\E_\xi[\xi_i]\norm{y_i}^2\notag\\
 &=p\sum_i\norm{y_i}^2=s\,t_y.
\label{eq:hausdorff-rademacher-energy}
\end{align}
Substituting \eqref{eq:hausdorff-rademacher-energy} into
\eqref{eq:hausdorff-rademacher-bound} gives
\begin{equation}\label{eq:hausdorff-first-moment}
 \E[\Delta]\le4\sqrt{t_y/s}.
\end{equation}

For the second moment, apply the Efron--Stein inequality
(\Cref{prop:efron-stein}). Let $\Delta^{(i)}$ be $\Delta$ with the
selector $\xi_i$ replaced by $\xi_i'$. Write
$$
 g(v):=\frac1n\sum_j|\ip v{y_j}|
       -\frac1s\sum_j\xi_j|\ip v{y_j}|,
$$
and let $g^{(i)}(v)$ be the same expression with $\xi_i$ replaced by
$\xi_i'$. By \eqref{eq:hausdorff-selector-form},
$\Delta=\sup_{\norm v\le1}|g(v)|$ and
$\Delta^{(i)}=\sup_{\norm v\le1}|g^{(i)}(v)|$.
For every $v$ with $\norm v\le1$, the triangle inequality gives
\begin{align*}
 |g(v)|
 &\le |g^{(i)}(v)|+|g(v)-g^{(i)}(v)|\\
 &\le\Delta^{(i)}+
       \sup_{\norm u\le1}|g(u)-g^{(i)}(u)|.
\end{align*}
Taking the supremum over $v$ and then exchanging $g$ and $g^{(i)}$
gives the first inequality below. Only the $i$-th summand differs
between these two functions, so
\begin{align}
 |\Delta-\Delta^{(i)}|
 &\le\sup_{\norm v\le1}|g(v)-g^{(i)}(v)|\notag\\
 &=\frac1s\sup_{\norm v\le1}
       |\xi_i-\xi_i'|\,|\ip v{y_i}|\notag\\
 &=\frac{|\xi_i-\xi_i'|}{s}\norm{y_i}.
\label{eq:hausdorff-coordinate-change}
\end{align}
The selectors $\xi_i,\xi_i'$ are independent $\mathrm{Bernoulli}(p)$
variables, so
\begin{align*}
 \E[(\xi_i-\xi_i')^2]
 &=\E[\xi_i^2]+\E[(\xi_i')^2]-2\E[\xi_i]\E[\xi_i']\\
 &=2p-2p^2=2p(1-p).
\end{align*}
Substituting this identity and \eqref{eq:hausdorff-coordinate-change}
into Efron--Stein and using $pn=s$ gives
\begin{align*}
 \Var\Delta
 &\le\frac12\sum_i\E[(\Delta-\Delta^{(i)})^2]\\
 &\le\frac1{2s^2}\sum_i\E[(\xi_i-\xi_i')^2]\norm{y_i}^2\\
 &=\frac{p(1-p)}{s^2}\sum_i\norm{y_i}^2
 =\frac{1-p}{s}t_y\le\frac{t_y}{s}.
\end{align*}
Together with \eqref{eq:hausdorff-first-moment}, this gives
$$
 \E[\Delta^2]=\Var\Delta+(\E[\Delta])^2
 \le\frac{t_y}{s}+\frac{16t_y}{s}=\frac{17t_y}{s}.
$$
\end{proof}

Now, we are ready to prove \Cref{lem:retained-sampling}.

\begin{proof}[Proof of \Cref{lem:retained-sampling}]
Recall the setup: we must bound $|\min_{K_Q}f-\min_K f|$, where
$K_Q$ and $K$ are the compact convex sets encoding the sampled and full
fractional cuts and $f(w)=w^TDw$ is the quadratic objective, with
$D=\diag(1/\lambda_j)$ and $|\lambda_j|>\tau$ for every retained $j$.
Set $\Delta=d_H(K_Q,K)$. We first bound the gap of the minima in
terms of $\Delta$, then apply \Cref{lem:hausdorff-moments} to take expectations.

The key point is that, although $D$ has operator norm
$\norm D_{\mathrm{op}}\le1/\tau$, which would naively make the gradient
of $f$ as large as $1/\tau$, the gradient is in fact bounded by two
on $K$. Since $f(w)=\sum_j w_j^2/\lambda_j$, differentiating
coordinate by coordinate gives
$$
 \frac{\partial f}{\partial w_j}(w)=\frac{2w_j}{\lambda_j}=2(Dw)_j,
 \qquad \nabla f(w)=2Dw.
$$
It therefore suffices to show that $\norm{Dw}\le1$ for $w\in K$.
For $w=\frac1n\sum_i x_iy_i\in K$,
$$
 (Dw)_j=\frac1n\sum_i x_i\phi_j(i)
        =\frac1{\sqrt n}\ip{x}{u_j}.
$$
Since the retained $u_j$ form an orthonormal family,
$$
 0\le\norm{x-\sum_{j\text{ retained}}\ip{x}{u_j}u_j}_2^2
 =\norm{x}_2^2-\sum_{j\text{ retained}}\ip{x}{u_j}^2.
$$
Thus $\sum_{j\text{ retained}}\ip{x}{u_j}^2\le\norm{x}_2^2$
(Bessel's inequality). Substituting this bound gives
\begin{align}
 \norm{Dw}^2
 &=\frac1n\sum_{j\text{ retained}}\ip{x}{u_j}^2\notag\\
 &\le\frac{\norm{x}_2^2}{n}
 =\frac1n\sum_i x_i^2\le1,
 \qquad \norm{\nabla f(w)}=2\norm{Dw}\le2.
\label{eq:retained-gradient}
\end{align}
For $w\in K$ and any $z$ with $\norm{z-w}\le\Delta$, the second-order
Taylor expansion of $f$ at $w$ is exact, since $f$ is quadratic.
Differentiating $\nabla f(w)=2Dw$ once more gives $\nabla^2f(w)=2D$, so
\begin{align}
 |f(z)-f(w)|
 &=\left|\ip{\nabla f(w)}{z-w}
       +\frac12(z-w)^T\nabla^2f(w)(z-w)\right|\notag\\
 &=\left|2\ip{Dw}{z-w}+(z-w)^TD(z-w)\right|\notag\\
 &\le2\norm{Dw}\norm{z-w}
       +\norm D_{\mathrm{op}}\norm{z-w}^2\notag\\
 &\le2\Delta+\Delta^2/\tau,
\label{eq:retained-objective-perturbation}
\end{align}
using \eqref{eq:retained-gradient} and $\norm D_{\mathrm{op}}\le1/\tau$.

Choose a minimizer $w_*$ of $f$ over $K$. Since
$d_H(K_Q,K)=\Delta$, some $z\in K_Q$ satisfies
$\norm{z-w_*}\le\Delta$, so \eqref{eq:retained-objective-perturbation} gives
$$
 \min_{K_Q}f\le f(z)\le f(w_*)+2\Delta+\Delta^2/\tau
 =\min_K f+2\Delta+\Delta^2/\tau.
$$
Conversely, choose a minimizer $z_*$ of $f$ over $K_Q$ and a point
$w\in K$ with $\norm{w-z_*}\le\Delta$; applying
\eqref{eq:retained-objective-perturbation} at this population point $w$ gives
$$
 f(z_*)\ge f(w)-2\Delta-\Delta^2/\tau
 \ge\min_K f-2\Delta-\Delta^2/\tau.
$$
Combining the two directions,
\begin{equation}\label{eq:retained-minima-perturbation}
 |\min_{K_Q}f-\min_K f|\le2\Delta+\Delta^2/\tau.
\end{equation}

To apply \Cref{lem:hausdorff-moments}, we now bound the average
squared length of our vectors:
\begin{align}
 t_y=\frac1n\sum_i\norm{y_i}^2
 &=\sum_j\lambda_j^2\left(\frac1n\sum_i\phi_j(i)^2\right)\notag\\
 &=\sum_j\lambda_j^2\le\norm A_F^2/n^2\le1,
\label{eq:retained-vector-energy}
\end{align}
where we used $\frac1n\sum_i\phi_j(i)^2=1$ for each retained $j$.
Taking expectations in \eqref{eq:retained-minima-perturbation} and
applying \Cref{lem:hausdorff-moments} gives
\begin{align*}
 \E\left[\left|\min_{K_Q}f-\min_K f\right|\right]
 &\le2\E[\Delta]+\frac{\E[\Delta^2]}{\tau}\\
 &\le8\sqrt{t_y/s}+\frac{17t_y}{\tau s}\\
 &\le\frac8{\sqrt s}+\frac{17}{\tau s}.
\end{align*}
The last inequality uses \eqref{eq:retained-vector-energy}.
By \eqref{eq:moment-objectives}, this proves \eqref{eq:mincompare}.
\end{proof}

\subsection{The sampling error for MAX-CUT}\label{subsec:upper-assembly}
\begin{proof}[Proof of \Cref{thm:sampling}]
It remains to bound the last term in
\eqref{eq:upper-error-decomposition} and the first term in
\eqref{eq:upper-sampling-error}. For the former, $A=L+R$ and
$\norm{x}_2^2\le n$ for $x\in[-1,1]^n$ give
\begin{align}
 \left|
 \frac{\min_x x^TLx}{n^2}-\frac{\min_x x^TAx}{n^2}
 \right|
 &\le\frac1{n^2}\sup_{x\in[-1,1]^n}|x^TRx|\notag\\
 &\le\frac{\norm R_{\mathrm{op}}}{n}\le\tau.
\label{eq:full-residual}
\end{align}

To bound the first term in \eqref{eq:upper-sampling-error}, let
$Z=\sum_{i<j}A_{ij}\xi_i\xi_j$. Since $A$ has zero diagonal and
$p=s/n$,
\begin{align}
 \one^TPAP\one&=2Z,\notag\\
 \E\left[\frac{2Z}{s^2}\right]
 &=\frac2{s^2}\sum_{i<j}A_{ij}\E[\xi_i\xi_j]
 =\frac{2p^2}{s^2}\sum_{i<j}A_{ij}
 =\frac{\one^TA\one}{n^2}.
\label{eq:density-mean}
\end{align}
For distinct $i,j,k$, independence and $\xi_i^2=\xi_i$ give
\begin{align*}
 \Var(\xi_i\xi_j)&=p^2-p^4\le p^2,\\
 \Cov(\xi_i\xi_j,\xi_i\xi_k)
 &=\E[\xi_i^2\xi_j\xi_k]
   -\E[\xi_i\xi_j]\E[\xi_i\xi_k]\\
 &=p^3-p^4\le p^3.
\end{align*}
Products associated with disjoint pairs are independent and have zero
covariance. There are at most $\binom n2$ edges and
$n\binom{n-1}{2}$ unordered pairs of edges sharing a vertex. Expanding
the variance of the sum therefore gives
\begin{equation}\label{eq:density-variance}
 \Var Z\le\binom n2p^2+2n\binom{n-1}{2}p^3
 \le\frac{n^2p^2}{2}+n^3p^3.
\end{equation}
Using \eqref{eq:density-mean}, Cauchy--Schwarz yields
\begin{align}
 \E\left[\left|\frac{\one^TPAP\one}{s^2}
       -\frac{\one^TA\one}{n^2}\right|\right]
 &=\frac2{s^2}\E[|Z-\E[Z]|]\nonumber\\
 &\le\frac2{s^2}\sqrt{\E[(Z-\E[Z])^2]}
 =\frac{2\sqrt{\Var Z}}{s^2}\nonumber\\
 &\le\sqrt{2/s^2+4/s}\le3/\sqrt s.
 \label{eq:density}
\end{align}
The last line uses \eqref{eq:density-variance}, $p=s/n$, and $s\ge1$.

We take expectations in \eqref{eq:upper-error-decomposition} and apply
\eqref{eq:full-residual} and
\Cref{lem:residual-sampling,lem:retained-sampling}. Substituting the
resulting bound and \eqref{eq:density} into
\eqref{eq:upper-sampling-error} gives
$$
 \E\left[\left|\frac{\MC(G[Q])}{s^2}-\frac{\MC(G)}{n^2}\right|\right]
 \le\frac14\left[
 (C_{\mathrm{res}}+1)\tau
 +\frac{C_{\mathrm{res}}+17}{\tau s}
 +\frac{C_{\mathrm{res}}+11}{\sqrt s}\right].
$$
Setting $\tau=1/\sqrt s\in(0,1]$ proves \Cref{thm:sampling} with
$C_0=(3C_{\mathrm{res}}+29)/4$.
\end{proof}

\begin{proof}[Proof of \Cref{thm:upper}]
The case $n\le s$ is trivial: querying the whole graph uses
$O(s^2)=O(1/\eps^4)$ pair queries and gives the exact MAX-CUT value.
Below we assume $n>s$ and analyze \Cref{alg:canonical}.
The estimator can fail only if one of two
events occurs: the sample has more than $2s$ vertices, causing an early
return to enforce the query budget, or the rescaled MAX-CUT value of
the sampled subgraph has error greater than $\eps n^2$. We bound the
probabilities of these two events separately.

First, $|Q|=\sum_i\xi_i$ is a sum of independent Bernoulli$(s/n)$
variables with mean $s$. Applying the Chernoff bound in \Cref{prop:chernoff} with $\mu=s$
and $\delta=1$ gives
$$
 \Pr(|Q|>2s)\le\exp\left(-\frac{1^2s}{2+1}\right)
 =e^{-s/3}\le e^{-C/3},
$$
since $s\ge C/\eps^2\ge C$.

Second, define the normalized sampling error
$Z=|\MC(G[Q])/s^2-\MC(G)/n^2|$.
By \Cref{thm:sampling}, $\E[Z]\le C_0/\sqrt s$. Applying Markov's
inequality at threshold $\eps$ gives
$$
 \Pr(Z>\eps)\le\frac{\E[Z]}{\eps}
 \le\frac{C_0}{\eps\sqrt s}
 \le\frac{C_0}{\sqrt C},
$$
using $s\ge C/\eps^2$.

When neither event occurs, the algorithm returns an estimate within
$\eps n^2$ of $\MC(G)$. A union bound therefore gives
$$
 \Pr(\text{estimator fails})
 \le\Pr(|Q|>2s)+\Pr(Z>\eps)
 \le e^{-C/3}+\frac{C_0}{\sqrt C}<\frac13
$$
for a sufficiently large constant $C$.

The tester makes $O(s^2)=O(1/\eps^4)$ non-adaptive pair queries.
Generating the sample by geometric gaps takes $O(s)$ operations,
and enumerating all cuts takes $O(2^{2s}s^2)=2^{O(1/\eps^2)}$ time.
\end{proof}


\section{Hard distributions}\label{sec:hard}

By Yao's minimax principle, our randomized lower bounds reduce to
showing that every deterministic algorithm making sufficiently few queries
cannot distinguish two hard
distributions with high probability.
We use the same hard distributions for all our lower bounds. The no distribution is a uniform distribution over all graphs. The yes distribution first chooses a uniformly random hidden partition and then samples edges independently, making crossing pairs connect slightly more likely and within-part pairs slightly less likely.

\begin{definition}[Hard distributions]\label{def:hard}
Fix an integer $n\ge2$ and $0<\eps\le1/160$, and set $\rho=40\eps$.
The \emph{yes distribution} $\mcald_1$ and the \emph{no distribution}
$\mcald_0$ are distributions on simple graphs with vertex set $[n]$,
defined as follows.

Under $\mcald_0$, each unordered pair is independently included as an
edge with probability $1/2$. Here and below we use $Y_{uv}\in\{-1,1\}$ to denote the edge indicator of $(u,v)\in\binom{[n]}{2}$, i.e., $(u,v)$ are connected in the graph if and only if $Y_{uv}=-1$.

To sample a graph from $\mcald_1$, first draw independent
uniform hidden signs $\Sigma_1,\ldots,\Sigma_n\in\{-1,1\}$.
Conditional on $\Sigma=(\Sigma_1,\ldots,\Sigma_n)$,
for each unordered pair $(u,v)$ we connect them with independent probability
\begin{equation}\label{eq:channel}
  \Pr(Y_{uv}=-1\mid\Sigma)
  =\frac{1-\rho \Sigma_u\Sigma_v}{2}.
\end{equation}
We write $\mcald_1^{\mathrm{joint}}$ for the joint distribution of
$\Sigma$ and the resulting graph $G$.

\end{definition}

Below, we show that the MAX-CUT values of graphs drawn from the two distributions are well separated with high probability.

\begin{lemma}\label{lem:gap}
For $0<\eps\le1/160$ and $n\ge\lceil1/\eps^2\rceil$, put
\begin{align*}
 {u_*}&=n^2/8+n^{3/2},\\
 {l_*}&=(1/2+20\eps)(n^2-9n)/4-n.
\end{align*}
Then ${l_*}-{u_*}>2\eps n^2$, and
$$
\begin{aligned}
 \Pr_{G\sim\mcald_0}(\MC(G)>{u_*})
 <0.001,\qquad
 \Pr_{G\sim\mcald_1}(\MC(G)<{l_*})
 <0.023.
\end{aligned}
$$
\end{lemma}
\begin{proof}
Fix a cut $\sigma\in\{-1,1\}^n$, and let $m_\sigma\le n^2/4$
be its number of crossing pairs. Under $\mcald_0$, its cut
size $X_\sigma=\cutg(\sigma)$ is a sum of $m_\sigma$ independent and uniform
random bits. Thus $\E [X_\sigma]=m_\sigma/2\le n^2/8$.
Hoeffding's inequality (\Cref{prop:hoeffding}) gives
$$
 \Pr_{\mcald_0}(X_\sigma>u_*)
 \le\Pr_{\mcald_0}(X_\sigma-\E [X_\sigma]>n^{3/2})
 \le\exp\left(-\frac{2n^3}{m_\sigma}\right)
 \le e^{-8n}.
$$
A union bound over the $2^n$ cuts gives
$$
 \Pr_{\mcald_0}(\MC(G)>u_*)
 \le\sum_{\sigma\in\{-1,1\}^n}\Pr_{\mcald_0}(X_\sigma>u_*)
 \le2^ne^{-8n}<0.001.
$$

Under $\mcald_1$, let $S=\sum_i\Sigma_i$.
By Hoeffding's inequality (\Cref{prop:hoeffding}), $S$ concentrates around $0$ with high probability:
$$
 \Pr(|S|>3\sqrt n)
 \le2\exp\left(-\frac{2(3\sqrt n)^2}{4n}\right)=2e^{-9/2}.
$$
Fix any hidden partition represented by $\sigma$ satisfying $|\sum_i\sigma_i|\le3\sqrt n$.
Its number of crossing pairs is
$m_\sigma=(n^2-(\sum_i\sigma_i)^2)/4\ge(n^2-9n)/4$.
Conditional on $\Sigma=\sigma$, its cut size $X=\cutg(\sigma)$
is a sum of $m_\sigma$ independent Bernoulli variables, each of mean
$1/2+20\eps$. Its mean $\mu_\sigma$ therefore satisfies
$\mu_\sigma\ge l_*+n$. The concentration bound (\Cref{prop:hoeffding}) gives,
$$
 \Pr(X<l_*\mid\Sigma=\sigma)
 \le\Pr(X-\mu_\sigma<-n\mid\Sigma=\sigma)
 \le\exp\left(-\frac{2n^2}{m_\sigma}\right)\le e^{-8}.
$$
Since $\MC(G)\ge\cutg(\Sigma)$, we have
$$
 \Pr_{\mcald_1}(\MC(G)<l_*)\le2e^{-9/2}+e^{-8}<0.023.
$$
Lastly, the two thresholds $l_*,u_*$ are far apart:

$$
 {l_*}-{u_*}=5\eps n^2-n^{3/2}-(17/8+45\eps)n>2\eps n^2.
$$
\end{proof}

\section{The non-adaptive lower bound}\label{sec:nonadaptive}
In this section, we give a simple proof of the non-adaptive query lower
bound for MAX-CUT. 

\begin{theorem}[Non-adaptive quartic lower bound]\label{thm:na}
For every parameter $0<\eps\le 0.005$ and integer $n\ge\lceil 1/\eps^2\rceil$,
any randomized non-adaptive tester for MAX-CUT with error parameter $\eps$ and success probability at least $2/3$ on every input
requires $\Omega(1/\eps^4)$ pair queries in the worst case.
\end{theorem}

Fix a deterministic non-adaptive algorithm and hence a query graph $H$.
The key observation is that one plus the $\chi^2$-divergence between the transcript distributions given yes and no inputs is exactly the high-temperature form of an Ising partition function on $H$. We subsequently bound it using Boolean hypercontractivity.

In this section, we use $\mcald_{1,H}$ and
$\mcald_{0,H}$ to denote the distributions of the answer vector
$(Y_{uv})_{(u,v)\in E(H)}$ under $\mcald_1$ and $\mcald_0$,
respectively.

We use the following standard Boolean hypercontractivity inequality.

\begin{lemma}[Boolean hypercontractivity {\citep[Theorem~9.21]{ODonnell2014}}]
\label{lem:hypercontractivity}
Fix integers $m\ge1$ and $d\ge0$, and let $f\colon\R^m\to\R$
be a multilinear polynomial of degree at most $d$.
For $X\sim\Unif_m$ and every real $p\ge2$,
$$
 \bigl(\E[|f(X)|^p]\bigr)^{1/p}
 \le (p-1)^{d/2}\bigl(\E[f(X)^2]\bigr)^{1/2}.
$$
Both expectations are over $X$.
\end{lemma}

\begin{lemma}\label{lem:fixeddesign}
Let $H$ be a fixed query graph, and write $q:=|E(H)|$. For independent uniform signs
$\Xi_1,\ldots,\Xi_n$,
\begin{equation}\label{eq:replica}
 1+\chi^2(\mcald_{1,H}\Vert \mcald_{0,H})
 =\E_\Xi\left[\prod_{(u,v)\in E(H)}(1+\rho^2\Xi_u\Xi_v)\right].
\end{equation}
If $a=e\rho^2\sqrt q<1$, then
$$
 \chi^2(\mcald_{1,H}\Vert \mcald_{0,H})\le\frac{a^2}{1-a}.
$$
\end{lemma}

The right-hand side of \eqref{eq:replica} is exactly a normalized partition function of an Ising model with coupling graph $H$.

\begin{proof}
Fix a query transcript $y\in\{-1,1\}^{E(H)}$. Under the no distribution,
its probability is $\mcald_{0,H}(y)=2^{-q}$. Under the yes distribution,
the queried answers are independent conditional on $\Sigma$, which is distributed as $\Unif_n$. Averaging over $\Sigma$ gives
$$
 \mcald_{1,H}(y)
 =2^{-q}\E_{\Sigma\sim\Unif_n}\left[
   \prod_{(u,v)\in E(H)}(1+\rho y_{uv}\Sigma_u\Sigma_v)\right].
$$
Thus the likelihood ratio is
$$
 \ell(y):=\frac{\mcald_{1,H}(y)}{\mcald_{0,H}(y)}
 =\E_{\Sigma\sim\Unif_n}\left[
   \prod_{(u,v)\in E(H)}(1+\rho y_{uv}\Sigma_u\Sigma_v)\right].
$$
Therefore,
\begin{align*}
 \E_{Y\sim\mcald_{0,H}}[\ell(Y)^2]
 &=\E_{\Sigma,\Sigma',Y}\left[
   \prod_{(u,v)\in E(H)}
   (1+\rho Y_{uv}\Sigma_u\Sigma_v)
   (1+\rho Y_{uv}\Sigma'_u\Sigma'_v)\right]\\
 &=\E_{\Sigma,\Sigma'}\left[
   \prod_{(u,v)\in E(H)}
   (1+\rho^2\Sigma_u\Sigma'_u\Sigma_v\Sigma'_v)\right]\\
 &=\E_{\Xi\sim\Unif_n}\left[
   \prod_{(u,v)\in E(H)}(1+\rho^2\Xi_u\Xi_v)\right],
\end{align*}
where $\Sigma'$ is an independent copy of $\Sigma$, and $\Xi_u=\Sigma_u\Sigma'_u$ is also uniform. This proves
\eqref{eq:replica}.

We now bound the expectation in \eqref{eq:replica} using an
exponential-moment estimate for sums of sign products.
Such exponential-moment estimates are standard; see, e.g.,
\citep[proof of Theorem~2]{cortinovis2022randomized}.
For completeness, we give a short derivation from Boolean
hypercontractivity.

Let
$$
 Z=\sum_{(u,v)\in E(H)}\Xi_u\Xi_v.
$$
Independence of $\Xi_1,\dots,\Xi_n$ gives $\E[Z]=0$ and $\E[Z^2]=q$.
Since $Z$ is a polynomial of degree at most two in the independent
uniform signs $\Xi_1,\ldots,\Xi_n$, \Cref{lem:hypercontractivity} gives
\begin{equation}\label{eq:query-quadratic-moments}
 \bigl(\E[|Z|^p]\bigr)^{1/p}\le(p-1)\sqrt q,
 \qquad p\ge2.
\end{equation}
Therefore, we can bound \eqref{eq:replica} as follows. Applying $1+x\le e^x$ and the Taylor expansion of $e^{\rho^2Z}$, we obtain
$$
\begin{aligned}
 1+\chi^2(\mcald_{1,H}\Vert\mcald_{0,H})
 &=\E_\Xi\left[\prod_{(u,v)\in E(H)}
       (1+\rho^2\Xi_u\Xi_v)\right]\\
 &\le\E_\Xi\left[\exp\left(\rho^2
       \sum_{(u,v)\in E(H)}\Xi_u\Xi_v\right)\right]\\
 &=1+\sum_{k\ge 2}\frac{\rho^{2k}\E[Z^k]}{k!}\\
 &\le1+\sum_{k\ge 2}\frac{\rho^{2k}\E[|Z|^k]}{k!}\\
 &\le1+\sum_{k\ge 2}\frac{\rho^{2k}(k-1)^k q^{k/2}}{k!}\\
 &\le1+\sum_{k\ge 2}a^k=1+\frac{a^2}{1-a}.
\end{aligned}
$$
The third inequality applies \eqref{eq:query-quadratic-moments} with $p=k$
to $\E[|Z|^k]$. The last inequality uses $k!\ge(k/e)^k$ and
$a=e\rho^2\sqrt q$, and the geometric series converges since $a<1$.
Subtracting one proves the bound.
\end{proof}

Now, we prove the nonadaptive quartic lower bound.
\begin{proof}[Proof of \Cref{thm:na}]
Set $c_*=1/(16e^2\cdot40^4)$ and consider algorithms that make at most $c_*/\eps^4$ pair queries. By \Cref{lem:gap}, the MAX-CUT values under
$\mcald_1$ and $\mcald_0$ are separated by more than $2\eps n^2$ with probability at least $0.976$.
Thus a tester with success probability at least $2/3$ on every
input would distinguish $\mcald_1$ from $\mcald_0$ by comparing
its estimate with $(l_*+u_*)/2$. When the two distributions are
chosen with equal probability, a union bound gives average
distinguishing success at least $2/3-0.024/2>5/8$.
By Yao's minimax principle, it suffices to show that every
deterministic non-adaptive algorithm making at most this many queries
has average distinguishing success below $5/8$ for these same
two distributions.

Fix any deterministic non-adaptive algorithm with this budget, and its query graph $H$ with
$q:=|E(H)|\le c_*/\eps^4$. Since $\rho=40\eps$, the parameter
in \Cref{lem:fixeddesign} satisfies
$$
 a=e\rho^2\sqrt q\le e\cdot40^2\sqrt{c_*}=\frac14.
$$
That lemma and \Cref{prop:tvd_chi2} therefore give
$$
 \TV(\mcald_{1,H},\mcald_{0,H})
 \le\frac12\sqrt{\frac{a^2}{1-a}}
 \le\frac{1}{4\sqrt3}<\frac16.
$$
Hence its distinguishing success when the two input distributions are equally likely is
at most $7/12<5/8$, proving the lower bound.
\end{proof}

\begin{remark}
It is well-known that any adaptive tester can be simulated by a non-adaptive tester with a quadratic blowup in the number of queries~\citep[Lemma~4.1]{GoldreichTrevisan2003}.

Our $\Omega(1/\eps^4)$ non-adaptive lower bound therefore gives only
an $\Omega(1/\eps^2)$ adaptive lower bound through this reduction.
Obtaining an $\omega(1/\eps^2)$ adaptive lower bound this way would
require an $\omega(1/\eps^4)$ non-adaptive lower bound, which our
$O(1/\eps^4)$ upper bound in \Cref{thm:upper} rules out.
\end{remark}

\begin{remark}
We cannot apply \Cref{lem:fixeddesign} to an adaptive algorithm simply
by conditioning on its final query graph, because that graph itself depends
on the answers. The equation \eqref{eq:replica} no longer holds simply because the query graph $H$ is not fixed and depends on the answers. In the next section, we
instead bound the information revealed by the adaptive transcript.
\end{remark}

\section{A cubic adaptive lower bound}\label{sec:warmup}

As noted above, the key challenge is the adaptivity of algorithms. We can no longer rewrite the $\chi^2$-divergence as a clean partition function. We overcome this challenge by bounding the KL divergence between the transcript distributions by $\rho^2$ times the expected sum of squared conditional expectations of the queried products $\Sigma_u\Sigma_v$, given the final transcript. For each fixed transcript, we can then bound the sum in terms of the information it reveals about the hidden partition. This bound holds for every set of queried pairs, including one chosen adaptively. Controlling the first two moments of that information gives the cubic lower bound.

\begin{theorem}[Cubic adaptive lower bound]\label{thm:cubic}
There exists $\eps_*>0$ such that for every $0<\eps\le\eps_*$ and every
integer $n\ge\lceil1/\eps^2\rceil$, any randomized adaptive tester for
MAX-CUT with error parameter $\eps$ and success probability at least $99/100$ requires $\Omega(1/\eps^3)$ pair
queries in the worst case.
\end{theorem}


By Yao's minimax principle and \Cref{lem:gap}, it suffices to show that every deterministic adaptive algorithm with a small enough query complexity cannot distinguish the yes and no distributions. More specifically, for a fixed algorithm, it suffices to show that the distributions of its query transcripts are indistinguishable given yes and no inputs, which we write as $\mcald_{1,T}$ and $\mcald_{0,T}$. We may assume that the deterministic algorithm always makes exactly $q$ distinct queries. Repeated queries reveal no new information and can be simulated using their earlier answers. If the algorithm stops early, it can query previously unqueried pairs while ignoring their answers. 

Write $Y_t=Y_{E_t}$ for the answer at step $t$, and define transcripts
$$
 T_t=((E_1,Y_1),\ldots,(E_t,Y_t)),\qquad T=T_q,
$$
with $T_0$ empty. At step $t$, the algorithm uses a fixed function $f_t$
such that $E_t=f_t(T_{t-1})=(U_t,V_t)$; at a realized history
$T_{t-1}=h$, this is the fixed pair $f_t(h)$.

The following lemma bounds the squared pair correlations of an arbitrary
distribution in terms of its divergence from the uniform distribution; it
will be applied to the conditional distribution of $\Sigma$
given the transcript.

\begin{lemma}\label{lem:leveltwo}
For any probability distribution $\mu$ on $\{-1,1\}^n$, let
$k=\KL(\mu\Vert\Unif_n)$. For any set $E$ of distinct unordered pairs, let $r_{uv}=\E_{\Sigma\sim\mu}[\Sigma_u\Sigma_v]$,
and put $s=\sum_{(u,v)\in E}r_{uv}^2$.
Then
$$
 s\le e^2\left(2\sqrt{\frac{k}{\log e}}+\frac{k}{\log e}\right)^2
 \le\frac{6e^2k}{\log e}+\frac{3e^2k^2}{(\log e)^2}.
$$
\end{lemma}

The second inequality absorbs the cross term so that the bound involves only $k$ and $k^2$, and avoids a fractional coefficient.

\begin{remark}\label{rmk:leveltwo}
$r_{uv}$ is constantly $0$ on a uniform distribution and $s$ measures how far $\mu$ deviates from it.
This lemma says that this measure is upper bounded by a linear combination of $k$ and $k^2$.
The $k^2$ term is necessary:
if $E=\binom{[n]}{2}$ and $\mu$ is uniform on the all-$(+1)$ and all-$(-1)$
vectors, then $r_{uv}=1$ for every pair, $k=n-1$, and
$s=\binom n2=\Theta(k^2)$.
\end{remark}

\begin{proof}
We construct a function $g(x)=\sum_{(u,v)\in E}r_{uv}x_ux_v$ for
$x\in\{-1,1\}^n$, so $ \E_{x\sim\Unif_n}[g(x)]=0$ and $\E_{x\sim\Unif_n}[g(x)^2]= \E_\mu[g]=s$. Our strategy is to raise $\E_\mu[\lambda g]$ to $\E_{\Unif_n}[e^{\lambda g}]$ using the entropy variational inequality \eqref{eq:duality}, which can be handled by the hypercontractivity inequality.

Specifically, since the pairs in $E$ are distinct,
\begin{align*}
 \E_{x\sim\Unif_n}[g(x)]
 &=\sum_{(u,v)\in E}r_{uv}\E_{x\sim\Unif_n}[x_ux_v]=0,\\
 \E_{x\sim\Unif_n}[g(x)^2]
 &=\sum_{(u,v),(a,b)\in E}r_{uv}r_{ab}
       \E_{x\sim\Unif_n}[x_ux_vx_ax_b]\\
 &=\sum_{(u,v)\in E}r_{uv}^2=s,\\
 \E_\mu[g]
 &=\sum_{(u,v)\in E}r_{uv}\E_\mu[\Sigma_u\Sigma_v]
 =\sum_{(u,v)\in E}r_{uv}^2=s.
\end{align*}

Under $\Unif_n$, the coordinates are independent uniform signs,
so the hypercontractivity (\Cref{lem:hypercontractivity}) with $d=2$ gives, for every integer $j\ge2$,
$$
 \E_{\Unif_n}[|g|^j]\le(j-1)^j s^{j/2}.
$$
Let $\lambda>0$ and $a=e\lambda\sqrt s<1$, whose values will be determined later. We expand $e^{\lambda g}$
and use $\E_{\Unif_n}[g]=0$ and the absolute-moment bound to obtain
$$
\begin{aligned}
 \E_{\Unif_n}[e^{\lambda g}]
 &\le1+\sum_{j\ge2}\frac{\lambda^j(j-1)^j s^{j/2}}{j!}\\
 &\le1+\sum_{j\ge2}a^j=1+\frac{a^2}{1-a},
\end{aligned}
$$
where we use $j!\ge (j/e)^j$.
Using $\log(1+x)\le(\log e)x$ gives
$$
 \log\E_{\Unif_n}[e^{\lambda g}]
 \le\log\left(1+\frac{a^2}{1-a}\right)
 \le(\log e)\frac{a^2}{1-a}.
$$
If $s=0$ there is nothing to prove; if $k=0$, then $\mu=\Unif_n$ and
$s=0$. Otherwise applying \eqref{eq:duality} to $\lambda g$, we obtain
\begin{equation*}
    \E_\mu[\lambda g]
    \le\frac{k+\log\E_{\Unif_n}[e^{\lambda g}]}{\log e}
    \le\frac{k}{\log e}+\frac{a^2}{1-a}.
\end{equation*}
Since $\E_\mu[g]=s$, we have

$$
    s=\E_\mu[\lambda g]/\lambda\le \frac{k}{(\log e)\lambda}+\frac{a^2}{(1-a)\lambda}=\frac{ek\sqrt s}{(\log e)a}+\frac{ea\sqrt s}{1-a},
$$
$$s\le \left(\frac{ek}{(\log e)a}+\frac{ea}{1-a}\right)^2.$$
We choose $a=\sqrt{k/(\log e)}/(1+\sqrt{k/(\log e)})$,
which minimizes the right-hand side, and obtain
$$s\le e^2\left(2\sqrt{\frac{k}{\log e}}+\frac{k}{\log e}\right)^2.$$

Finally,
$$
 \left(6+\frac{3k}{\log e}\right)
 -\left(2+\sqrt{\frac{k}{\log e}}\right)^2
 =2\left(\sqrt{\frac{k}{\log e}}-1\right)^2\ge0,
$$
so the second inequality in the statement follows.
\end{proof}

Recall from \Cref{sec:hard} that $\mcald_1^{\mathrm{joint}}$ is the joint yes
distribution of $\Sigma$ and the graph.
We now define the prediction variables used below. For each
$1\le t\le q$, define
$$
 X_t=\Sigma_{U_t}\Sigma_{V_t},\qquad
 C_t=\E_{\mcald_1^{\mathrm{joint}}}[X_t\mid T_{t-1}],\qquad
 R_t=\E_{\mcald_1^{\mathrm{joint}}}[X_t\mid T_q],
$$
where the queried indices $U_t,V_t$ are random, being functions of the
transcript $T_{t-1}$. Thus $C_t$ is the prediction of $X_t$ from the transcript so
far, and $R_t$ is the prediction from the complete transcript. Since
$T_{t-1}$ is determined by $T_q$, the law of iterated expectations gives
$$
 C_t=\E_{\mcald_1^{\mathrm{joint}}}[R_t\mid T_{t-1}],
$$
so the current prediction is the expected value of the terminal
prediction given the transcript so far.

For $x\in(-1,1)$, let $d(x)$ denote the KL divergence of a $\{\pm1\}$ variable
with mean $x$ from a fair one. By definition we have
$$
 d(x)=\tfrac{1+x}{2}\log(1+x)+\tfrac{1-x}{2}\log(1-x).
$$
Expanding the logarithms, for $|x|\le1/2$ we obtain
\begin{equation}\label{eq:d}
 d(x)=(\log e)\sum_{j\ge1}\frac{x^{2j}}{(2j-1)(2j)}
 \le\frac{\log e}{2}\sum_{j\ge1}x^{2j}
 =\frac{(\log e)x^2}{2(1-x^2)}\le\frac23(\log e)x^2.
\end{equation}

To apply the pointwise bound of \Cref{lem:leveltwo} at a random complete
transcript $T$,
we need the first two moments of $\KL(\mu_T\Vert\Unif_n)$, where $\mu_T$
is the conditional distribution of $\Sigma$ given $T$.
Information density provides that control.

\begin{lemma}\label{lem:infodensity}
For every complete transcript $h$ with
$\Pr_{\mcald_1^{\mathrm{joint}}}(T=h)>0$, define
$$
 \mu_h(s)=\Pr_{\mcald_1^{\mathrm{joint}}}(\Sigma=s\mid T=h)\quad(s\in\{-1,1\}^n),
 \qquad k(h)=\KL(\mu_h\Vert\Unif_n).
$$
Set $K=k(T)$. Then
\begin{align}
 \E_{\mcald_1^{\mathrm{joint}}}[K]&\le\tfrac23(\log e)q\rho^2,\label{eq:Kfirst}\\
 \E_{\mcald_1^{\mathrm{joint}}}[K^2]&\le(\log e)^2\left(4\rho\sqrt q+\tfrac{16}{3}q\rho^2\right)^2.
 \label{eq:Ksecond}
\end{align}
\end{lemma}
These bounds are on the average over the random transcript, not
pointwise in $h$.
\begin{proof}
The hidden partition $\Sigma$ has marginal distribution $\Unif_n$, so the definition of
mutual information gives
\begin{align*}
 \E_{\mcald_1^{\mathrm{joint}}}[K]
 &=\sum_h\Pr(T=h)\sum_s\mu_h(s)\log\frac{\mu_h(s)}{2^{-n}}\\
 &=\sum_{h,s}\Pr(T=h,\Sigma=s)
      \log\frac{\Pr(T=h,\Sigma=s)}{\Pr(T=h)\Pr(\Sigma=s)}\\
 &=I(\Sigma;T).
\end{align*}
All probabilities here are under $\mcald_1^{\mathrm{joint}}$.
By the chain rule, since each queried address $E_t=f_t(T_{t-1})$ is
determined by the history,
$$
 I(\Sigma;T)=\sum_t I(\Sigma;Y_t\mid T_{t-1}).
$$
Fix a step $t$. Given $(\Sigma,T_{t-1})$ the answer $Y_t$ has mean
$\rho X_t$ with $|X_t|=1$, while given $T_{t-1}$ alone it has mean
$\rho C_t$; a $\{\pm1\}$ distribution of mean $m$ has entropy
$1-d(m)$. Hence
$$
\begin{aligned}
 I(\Sigma;Y_t\mid T_{t-1})
 &=H(Y_t\mid T_{t-1})-H(Y_t\mid\Sigma,T_{t-1})\\
 &=\E [1-d(\rho C_t)]-\bigl(1-d(\rho)\bigr)\\
 &=d(\rho)-\E [d(\rho C_t)].
\end{aligned}
$$
Summing over $t$ and applying \eqref{eq:d},
$$
 I(\Sigma;T)=q d(\rho)-\sum_t\E [d(\rho C_t)]
 \le q d(\rho)\le\tfrac23(\log e)q\rho^2.
$$
This proves \eqref{eq:Kfirst}, since
$\E_{\mcald_1^{\mathrm{joint}}} [K]=I(\Sigma;T)$.

For the second moment, the mutual-information chain rule controls only
$\E[K]$; to control the fluctuations of $K$ we pass to the
log-likelihood ratio $J=\sum_t J_t$, a sum over steps whose conditional
mean is $K=\E[J\mid T_q]$, so conditional Jensen reduces $\E[K^2]$ to
$\E[J^2]$. We decompose
$J=M_q+B_q$ into a martingale part $M_q$ and a drift part $B_q$:
orthogonality of martingale increments controls $M_q$ via the pointwise
bound $|J_t|\le2(\log e)\rho/(1-\rho)$, while a two-point divergence computation
controls $B_q$ term by term.

Below we give details. We define
$$
 J_t=\log\frac{1+\rho Y_tX_t}{1+\rho Y_tC_t},\qquad
 J=\sum_tJ_t.
$$
To see the meaning of this ratio, fix a reached history $T_{t-1}=h$.
The queried pair $(u,v)=f_t(h)$ is then fixed. Under
$\mcald_1^{\mathrm{joint}}$, the answer distribution in \eqref{eq:channel} gives,
for each $y\in\{-1,1\}$,
\begin{equation}\label{eq:answer-given-labels}
 \Pr(Y_t=y\mid\Sigma,T_{t-1}=h)
 =\frac{1+\rho y\Sigma_u\Sigma_v}{2}.
\end{equation}
If only the history is given, we average over the conditional distribution
of $\Sigma$:

\begin{equation}\label{eq:answer-given-history}
\begin{aligned}
 \Pr(Y_t=y\mid T_{t-1}=h)
 &=\E\left[\frac{1+\rho y\Sigma_u\Sigma_v}{2}
       \mathrel{\Big|} T_{t-1}=h\right]\\
 &=\frac{1+\rho y C_t(h)}{2}.
\end{aligned}
\end{equation}
Evaluating these probabilities at the realized history and answer
$y=Y_t$, their ratio is
$$
 \frac{(1+\rho Y_tX_t)/2}{(1+\rho Y_tC_t)/2}
 =\frac{1+\rho Y_tX_t}{1+\rho Y_tC_t}.
$$
Thus $J_t$ is the logarithm of the probability of the observed answer given
$(\Sigma,T_{t-1})$ divided by its probability given $T_{t-1}$ alone.
By the chain rule for probabilities, their sum $J$ is exactly the log-likelihood ratio of the complete transcript given $\Sigma$ against its marginal distribution. To see this, by \eqref{eq:answer-given-labels} and \eqref{eq:answer-given-history}, we obtain
\begin{align*}
      \log\frac{\mu_h(s)}{2^{-n}}=&\log \frac{\Pr(T=h\mid\Sigma=s)}{\Pr(T=h)}\\
      =&\log \prod_{t=1}^q\frac{1+\rho y_ts_{u_t}s_{v_t}}{1+\rho y_t C_t(h_{t-1})}\\
      =&\sum_{t=1}^q\log\frac{1+\rho y_ts_{u_t}s_{v_t}}{1+\rho y_t C_t(h_{t-1})}.
\end{align*}
Evaluating this identity at $(T,\Sigma)$ gives $J=\log\frac{\mu_T(\Sigma)}{2^{-n}}$.

By the definition of KL divergence,
\begin{equation}\label{eq:conditional-information-density}
 \E[J\mid T_q]
 =\sum_s\mu_{T_q}(s)\log\frac{\mu_{T_q}(s)}{2^{-n}}
 =\KL(\mu_{T_q}\Vert\Unif_n)=K.
\end{equation}
Conditioning now on $\Sigma$ together with the transcript prefix
$T_{t-1}$, put $A_t=\E[J_t\mid\Sigma,T_{t-1}]$ and write
$$
 J=M_q+B_q,\qquad M_q=\sum_t(J_t-A_t),\quad B_q=\sum_t A_t.
$$
The random increments $J_t$ and their conditional means $A_t$
are taken under $\mcald_1^{\mathrm{joint}}$. Conditioned on the hidden partition and the current transcript, the partial sums
$M_t=\sum_{j=1}^{t}(J_j-A_j)$ form a martingale, i.e.,
$\E[M_t\mid \Sigma, T_{t-1}]=M_{t-1}$.
We bound the increments. Write $J_t=f(X_t)-f(C_t)$ with
$f(z)=\log(1+\rho Y_tz)$; for $|z|\le1$,
$$
 |f'(z)|=\frac{(\log e)\rho}{|1+\rho Y_tz|}
 \le\frac{(\log e)\rho}{1-\rho}.
$$
The mean value theorem therefore gives
$$
 |J_t|=|f(X_t)-f(C_t)|
 \le\frac{(\log e)\rho}{1-\rho}|X_t-C_t|
 \le\frac{2(\log e)\rho}{1-\rho}.
$$
Since
$\E[J_t-A_t\mid \Sigma, T_{t-1}]=0$, applying
\Cref{prop:martingale-orthogonality} and the definition of conditional variance gives
\begin{equation}\label{eq:information-martingale-moment}
\begin{aligned}
 \E[M_q^2]
 &=\sum_{t=1}^q\E[(J_t-A_t)^2]\\
 &=\sum_{t=1}^q\E[\Var(J_t\mid\Sigma, T_{t-1})]\\
 &\le\sum_{t=1}^q\E[J_t^2]
 \le\frac{4(\log e)^2q\rho^2}{(1-\rho)^2}.
\end{aligned}
\end{equation}
Thus $\norm{M_q}_2\le2(\log e)\rho\sqrt q/(1-\rho)$, where
$\norm X_2=(\E[X^2])^{1/2}$ denotes the $L^2$ norm.

It remains to bound the drift $B_q=\sum_t A_t$. Conditional on
$(\Sigma,T_{t-1})$, the answer $Y_t\in\{\pm1\}$ has
mean $\rho X_t$, whereas its marginal distribution given $T_{t-1}$ alone has mean
$\rho C_t$. By definition, the drift $A_t=\E[J_t\mid\Sigma,T_{t-1}]$ is exactly the KL-divergence between these two distributions. 

We bound their $\chi^2$-divergence instead and transfer it to KL-divergence using \eqref{eq:klchi2}. If we use $\mcald_a,\mcald_b$ to denote distributions over $\{\pm1\}$ of mean $a,b$, then
\begin{align*}
 \chi^2(\mcald_a\Vert\mcald_b)
 &=\sum_{y\in\{-1,1\}}
    \frac{\bigl((1+ay)/2-(1+by)/2\bigr)^2}{(1+by)/2}\\
 &=\frac{(a-b)^2}{2}\left(\frac1{1+b}+\frac1{1-b}\right)
 =\frac{(a-b)^2}{1-b^2}.
\end{align*}
Thus \eqref{eq:klchi2} gives
$$
 0\le A_t\le(\log e)\frac{(\rho X_t-\rho C_t)^2}{1-\rho^2C_t^2}
 \le\frac{4(\log e)\rho^2}{1-\rho^2},
$$
using $|X_t-C_t|\le2$ and $|C_t|\le1$. Consequently,
\begin{equation}\label{eq:information-drift-bound}
 0\le B_q=\sum_tA_t\le\frac{4(\log e)q\rho^2}{1-\rho^2},
 \qquad
 \norm{B_q}_2=\sqrt{\E[B_q^2]}\le\frac{4(\log e)q\rho^2}{1-\rho^2}.
\end{equation}

Combining \eqref{eq:information-martingale-moment} and
\eqref{eq:information-drift-bound},
\begin{equation}\label{eq:information-density-moment}
 \norm J_2\le\norm{M_q}_2+\norm{B_q}_2
 \le(\log e)\left(\frac{2\rho\sqrt q}{1-\rho}
 +\frac{4q\rho^2}{1-\rho^2}\right).
\end{equation}
By \eqref{eq:conditional-information-density}, conditional Jensen gives
$$
 \E[K^2]=\E\left[(\E[J\mid T_q])^2\right]
 \le\E\left[\E[J^2\mid T_q]\right]=\E[J^2].
$$
Using $\rho\le1/2$ in the two denominators of
\eqref{eq:information-density-moment}, we conclude that
\begin{align*}
 \sqrt{\E[K^2]}
 &\le\norm J_2
 \le(\log e)\left(\frac{2\rho\sqrt q}{1-\rho}
 +\frac{4q\rho^2}{1-\rho^2}\right)\\
 &\le(\log e)\left(4\rho\sqrt q+\frac{16}{3}q\rho^2\right).
\end{align*}
Squaring both sides proves \eqref{eq:Ksecond}.
\end{proof}

\begin{lemma}[Cubic transcript bound]\label{lem:cubictranscript}
For every deterministic adaptive algorithm that always makes exactly $q$
distinct queries, let $\mcald_{1,T},\mcald_{0,T}$ respectively denote its query transcript distributions given yes and no inputs. Then
\begin{equation}\label{eq:cubic}
 \KL(\mcald_{1,T}\Vert \mcald_{0,T})\le(\log e)e^2\left(
 \frac{104}{3}q\rho^4+\frac{256}{3}q^{3/2}\rho^5
 +\frac{512}{9}q^2\rho^6\right).
\end{equation}
\end{lemma}
\begin{proof}
We bound the divergence by reducing it step by step. The KL chain rule
decomposes it over the answers. The queried address $E_t=f_t(T_{t-1})$
is the same deterministic function of the history under both input
distributions, so it contributes nothing. Given $T_{t-1}$, the answer under the no distribution is uniform, while the answer under the yes distribution has mean $\rho C_t$. Hence, by \eqref{eq:d},
\begin{equation}\label{eq:cubic-predictive-kl}
 \KL(\mcald_{1,T}\Vert\mcald_{0,T})
 =\sum_{t=1}^q\E_{\mcald_1^{\textnormal{joint}}}[d(\rho C_t)]
 \le\tfrac23(\log e)\rho^2\sum_t\E[C_t^2].
\end{equation}
Next, recall that $C_t=\E[R_t\mid T_{t-1}]$.
Conditional Jensen gives
\begin{align}
 \sum_t\E[C_t^2]=\sum_t\E\left[(\E[R_t\mid T_{t-1}])^2\right]\le\sum_t\E\left[\E[R_t^2\mid T_{t-1}]\right]
 =\sum_t\E[R_t^2].
\label{eq:cubic-prediction-comparison}
\end{align}
So it remains to bound $\sum_t\E[R_t^2]$. Write $S=\sum_t R_t^2$. For a
fixed transcript $h$, let $\mu_h$ be the conditional distribution of $\Sigma$ given $T=h$; then the prediction at step $t$ is
$R_t(h)=\E_{\mu_h}[\Sigma_u\Sigma_v]$, where $(u,v)=E_t(h)$ is the
queried pair, so
$$
 S(h)=\sum_t R_t(h)^2
 =\sum_{(u,v)\in E(h)}\bigl(\E_{\mu_h}[\Sigma_u\Sigma_v]\bigr)^2.
$$
By \Cref{lem:leveltwo}, we get
$$
 S(h)\le\frac{6e^2k(h)}{\log e}
 +\frac{3e^2k(h)^2}{(\log e)^2},
 \qquad k(h)=\KL(\mu_h\Vert\Unif_n),
$$
for every $h$; in random-variable notation,
$S\le6e^2K/(\log e)+3e^2K^2/(\log e)^2$. Taking
expectations and applying \Cref{lem:infodensity},
\begin{equation}\label{eq:cubic-terminal-energy}
\begin{aligned}
 \E[S]
 &\le\frac{6e^2}{\log e}\E[K]
    +\frac{3e^2}{(\log e)^2}\E[K^2]\\
 &\le6e^2\cdot\tfrac23q\rho^2
    +3e^2\left(4\rho\sqrt q+\tfrac{16}{3}q\rho^2\right)^2\\
 &=e^2\left(52q\rho^2+128q^{3/2}\rho^3+\tfrac{256}{3}q^2\rho^4\right).
\end{aligned}
\end{equation}
Combining \eqref{eq:cubic-predictive-kl} and
\eqref{eq:cubic-prediction-comparison} gives
$\KL(\mcald_{1,T}\Vert\mcald_{0,T})
\le\tfrac23(\log e)\rho^2\E[S]$.
Substituting \eqref{eq:cubic-terminal-energy} proves \eqref{eq:cubic}.
\end{proof}

\begin{proof}[Proof of \Cref{thm:cubic}]
By Yao's minimax principle and \Cref{lem:gap}, it suffices to show that every deterministic algorithm that distinguishes $\mcald_1$ from $\mcald_0$ with average distinguishing success at least $9/10$ must make $\Omega(1/\varepsilon^3)$ queries in the worst case. Indeed, a tester with success probability at least $99/100$ on every graph would distinguish the two hard distributions with average success at least $99/100-0.024/2>9/10$.

We can assume without loss of generality that the algorithm always makes exactly $q$ distinct queries. Any algorithm with worst-case query
complexity at most $q$ converts into one of this form without changing
its output by querying unused pairs after its original stopping time.

Choose a sufficiently small absolute constant $a>0$. If
$q\le a/\rho^3$, then \Cref{lem:cubictranscript} gives
\begin{align*}
 \KL(\mcald_{1,T}\Vert\mcald_{0,T})
 \le(\log e)e^2\left(\frac{104}{3}a\rho
       +\frac{256}{3}a^{3/2}\rho^{1/2}
       +\frac{512}{9}a^2\right)<\frac{\log e}{8}
\end{align*}
for sufficiently small $\rho$. By \Cref{prop:tvd_chi2},
$$
 \TV(\mcald_{1,T},\mcald_{0,T})
 \le\sqrt{\frac{\KL(\mcald_{1,T}\Vert\mcald_{0,T})}{2\log e}}
 <\frac14.
$$
Every deterministic distinguishing rule therefore has average success
at most $(1+1/4)/2=5/8<9/10$. Hence the query complexity must be
$\Omega(1/\rho^3)=\Omega(1/\eps^3)$.
\end{proof}

\begin{remark}[Why this argument stops at the cubic bound]\label{rmk:entropy}
As we noted in \Cref{rmk:leveltwo}, the quadratic $k^2$ term of \Cref{lem:leveltwo} is genuinely necessary:
when $\mu$ is uniform on the all-$(+1)$ and all-$(-1)$ assignments, we have $s=\Theta(n^2)$ and $k=\Theta(n)$,
so $s=\Theta(k^2)$. This term produces the $q^2\rho^6$ term
in \eqref{eq:cubic}. The above argument therefore bottoms out at
$q\asymp1/\rho^3$.

However, \Cref{lem:leveltwo} treats $\mu$ as an arbitrary distribution, whereas the conditional distribution $\mu_T$ in our proof arises from a sequence of noisy pair queries. Improving the bound would require exploiting the structure of this conditional distribution.
\end{remark}

\section*{Acknowledgements}\label{sec:acknowledgements}
\addcontentsline{toc}{section}{Acknowledgements}

The author would like to thank Periklis~A.~Papakonstantinou for valuable suggestions in improving the presentation of an earlier version of this paper. The author would also like to thank Yucheng~Fu for helpful discussions.

\paragraph{Use of AI.}\label{sec:use-of-ai}

The author used ChatGPT to generate the technical proofs in this paper and to assist in drafting the manuscript. The author subsequently substantially rewrote the manuscript throughout and verified the correctness and originality of all its contents, including the references. The author takes full responsibility for the paper.

\newpage
\printbibliography
\clearpage
\appendix
\section{The MAX-2CSP upper bound}\label{app:max2csp}

We give a tight non-adaptive tester for MAX-2CSP of query complexity $O(1/\eps^4)$, which follows from a similar analysis as in \Cref{sec:upper}.


Write $F$ for an instance, $\mathrm{OPT}(F)$ for its optimum, and
$F[Q]$ for its induced instance on $Q$. The constants $C_0$ and
$C_{\mathrm{res}}$ below may differ from those in \Cref{sec:upper}.

\begin{theorem}[Quartic query upper bound for MAX-2CSP]
\label{thm:csp-upper}
For every $n\ge2$ and $0<\eps\le1$, there is a randomized
non-adaptive algorithm that estimates the MAX-2CSP value to within
additive error $\eps n^2$ with probability at least $2/3$ and makes
$O(1/\eps^4)$ pair queries in the worst case. Its running time is
$2^{O(1/\eps^2)}$ in the RAM model.
\end{theorem}

The algorithm samples variables, queries the induced instance, and
rescales its optimum, exactly as in \Cref{alg:canonical}.
\begin{center}
\begin{minipage}{0.94\linewidth}
\raggedright
\hrule\smallskip
\refstepcounter{algo}\label{alg:csp-canonical}%
\textbf{Algorithm \thealgo} Canonical tester for \Cref{thm:csp-upper}.
\begin{itemize}[leftmargin=1.5em,itemsep=2pt,topsep=4pt,parsep=0pt,partopsep=0pt]
 \item Set $s=\lceil C/\eps^2\rceil$ for a sufficiently large absolute constant $C$.
 \item If $n\le s$, query all $\binom n2$ pairs and return $\mathrm{OPT}(F)$.
 \item Otherwise include each variable in $Q$ independently with probability $s/n$.
 \begin{itemize}[leftmargin=1.2em,itemsep=2pt,topsep=2pt,parsep=0pt,partopsep=0pt]
  \item If $|Q|>2s$, return zero without making queries.
  \item Otherwise query all pairs in $Q$ and return
  $n^2\mathrm{OPT}(F[Q])/s^2$.
 \end{itemize}
\end{itemize}
\smallskip\hrule
\end{minipage}
\end{center}

We show that the error of the estimation is small in expectation.
\begin{theorem}\label{thm:csp-sampling}
There is an absolute constant $C_0>0$ such that the
following holds. Let $1\le s\le n$, and include each variable in $Q$
independently with probability $s/n$. Then
$$
 \E\left[\left|
 \frac{\mathrm{OPT}(F[Q])}{s^2}
 -\frac{\mathrm{OPT}(F)}{n^2}\right|\right]
 \le\frac{C_0}{\sqrt s}.
$$
\end{theorem}

Fix the instance $F$. Choose an ordering of the variables in each
constraint, and for $a,b\in\{0,1\}$ let $A^{(a,b)}_{ij}$ count the
constraints assigned to the ordered pair $(i,j)$ that are satisfied
when $x_i=a$ and $x_j=b$. Since the constraints are distinct, each
pair carries at most $B:=16$ Boolean constraints. Each constraint is
assigned to one ordered pair, so $0\le A^{(a,b)}_{ij}\le B$ and
$A^{(a,b)}_{ii}=0$.
Construct a symmetric $2n\times2n$ matrix $M$, indexed by $(i,a)$, by
$$
 M_{(i,a),(j,b)}
 =\frac{A^{(a,b)}_{ij}+A^{(b,a)}_{ji}}2
 \quad(i\ne j),\qquad
 M_{(i,a),(i,b)}=0.
$$
For an assignment $x$, set $z_{i,a}=\boldsymbol{1}\{x_i=a\}$.
The number of satisfied constraints is exactly
$$
 \sum_{i,j}\sum_{a,b}A^{(a,b)}_{ij}z_{i,a}z_{j,b}=z^TMz.
$$
We relax $z$ to be real numbers.
$$
 \mathcal X=\{z\in[0,1]^{[n]\times\{0,1\}}:
                  z_{i,0}+z_{i,1}=1\text{ for every }i\}.
$$
This relaxation does not change its maximum. Indeed, when all other variables
are fixed, the maximum is still linear in $z_{i,0}$ and $z_{i,1}$, which is
maximized at $(z_{i,0},z_{i,1})=(1,0)$ or $(0,1)$.

For the sampling analysis, set $p=s/n$ and let
$\xi_i\sim\mathrm{Bernoulli}(p)$ independently. Write
$Q=\{i:\xi_i=1\}$, and let $P$ be the $2n\times2n$ diagonal matrix
with $P_{(i,a),(i,a)}=\xi_i$ for both $a\in\{0,1\}$.
All expectations below are over the variable selectors unless another
source of randomness is specified.

The entries of $M$ have absolute value at most $B$, and each row has
at most $2(n-1)$ nonzero entries. Hence, for every index
$\alpha=(i,a)$,
\begin{equation}\label{eq:csp-matrix-energy}
 (M^2)_{\alpha\alpha}
 =\sum_{\gamma}M_{\alpha\gamma}^2\le2B^2n,
 \qquad \norm M_F^2\le4B^2n^2.
\end{equation}
Let $\theta_j$ and $u_j$ be the eigenvalues and orthonormal
eigenvectors of $M$. For a parameter $0<\tau\le1$ to be determined later, write
$$
 L=\sum_{|\theta_j|>\tau n}\theta_ju_ju_j^T,\qquad R=M-L.
$$
Thus $\norm R_{\mathrm{op}}\le\tau n$ and
$\norm R_F\le\norm M_F\le2Bn$. The matrices $L$ and $R$ are
deterministic because the instance is fixed.

The triangle inequality gives
\begin{align}
 \left|\frac{\mathrm{OPT}(F[Q])}{s^2}
       -\frac{\mathrm{OPT}(F)}{n^2}\right|
 &\le\frac1{s^2}\sup_{z\in\mathcal X}
       |z^TPRPz|\notag\\
 &\quad+\left|
   \max_{z\in\mathcal X}\frac{z^TPLPz}{s^2}
   -\max_{z\in\mathcal X}\frac{z^TLz}{n^2}
   \right|\notag\\
 &\quad+\frac1{n^2}\sup_{z\in\mathcal X}|z^TRz|.
 \label{eq:csp-error-decomposition}
\end{align}
We note that after spectral truncation, fixing all other variables need not make objectives such as $z^TLz$ linear in each $z_{i,0}$ under $z_{i,0}+z_{i,1}=1$. Hence we can no longer assume that the maxima are attained at integral points of $\mathcal X$.
The last term is at most $\tau$: since
$z_{i,0}^2+z_{i,1}^2\le(z_{i,0}+z_{i,1})^2=1$,
\begin{equation}\label{eq:csp-full-residual}
 \frac1{n^2}\sup_{z\in\mathcal X}|z^TRz|
 \le\frac{\norm R_{\mathrm{op}}}{n^2}
      \sup_{z\in\mathcal X}\norm{z}_2^2
 \le\tau.
\end{equation}
We bound the other two terms separately.

\subsection{The cost of discarding $R$ in the sample}
\begin{lemma}[Sampling the MAX-2CSP residual]\label{lem:csp-residual}
There is an absolute constant $C_{\mathrm{res}}>0$ such that
$$
 \frac1{s^2}\E\left[
 \sup_{z\in\mathcal X}|z^TPRPz|\right]
 \le C_{\mathrm{res}}
 \left(\tau+\frac1{\tau s}+\frac1{\sqrt s}\right).
$$
\end{lemma}
\begin{proof}
For $a,b\in\{0,1\}$, let $R^{(a,b)}$ be the $n\times n$ block
whose $(i,j)$ entry is $R_{(i,a),(j,b)}$, and put
$z^{(a)}=(z_{i,a})_{i=1}^n$ and
$P_{\mathrm{var}}=\diag(\xi_1,\ldots,\xi_n)$.
Since $P_{\mathrm{var}}$ acts on the variable index
and $0\le z^{(a)}_i\le1$,
\begin{align}
 z^TPRPz
 &=\sum_{a,b\in\{0,1\}}(z^{(a)})^TP_{\mathrm{var}}R^{(a,b)}P_{\mathrm{var}}z^{(b)},\notag\\
 \sup_{z\in\mathcal X}|z^TPRPz|
 &\le\sum_{a,b\in\{0,1\}}\norm{P_{\mathrm{var}}R^{(a,b)}P_{\mathrm{var}}}_{\infty\to1}.
 \label{eq:csp-residual-blocks}
\end{align}
The matrices $R^{(a,b)}$ are deterministic, and the selectors $\xi_i$
are independent across variable indices. Thus \Cref{lem:restriction}
applies to each block.

We bound the three terms in \Cref{lem:restriction}: the diagonal sum,
the $\ell_\infty\to\ell_1$ norm after removing the diagonal, and the
column norms of the matrix and its transpose. First,
$M_{(i,a),(i,b)}=0$, so
$R^{(a,b)}_{ii}=-L_{(i,a),(i,b)}$. For
$\alpha=(i,a)$ and $\gamma=(i,b)$, the cutoff on the retained
eigenvalues and Cauchy--Schwarz give
\begin{align}
 |R^{(a,b)}_{ii}|
 &\le\sum_{j\text{ retained}}|\theta_j|
       |u_j(\alpha)u_j(\gamma)|\notag\\
 &\le\frac1{\tau n}
 \left(\sum_j\theta_j^2u_j(\alpha)^2\right)^{1/2}
 \left(\sum_j\theta_j^2u_j(\gamma)^2\right)^{1/2}\notag\\
 &=\frac{\sqrt{(M^2)_{\alpha\alpha}
                       (M^2)_{\gamma\gamma}}}{\tau n}
 \le\frac{2B^2}{\tau}.
 \label{eq:csp-block-diagonal}
\end{align}
The equality uses $M^2=\sum_j\theta_j^2u_ju_j^T$; the last
inequality follows from \eqref{eq:csp-matrix-energy}.
Each block is a compression of $R$, so
$\norm{R^{(a,b)}}_{\mathrm{op}}\le\tau n$ and
$\norm{R^{(a,b)}}_F\le2Bn$. If
$H^{(a,b)}=R^{(a,b)}-\diag(R^{(a,b)})$, then
\begin{align}
 \norm{H^{(a,b)}}_{\infty\to1}
 &\le n\norm{R^{(a,b)}}_{\mathrm{op}}
       +\sum_i|R^{(a,b)}_{ii}|
 \le\tau n^2+\frac{2B^2n}{\tau},\notag\\
 \norm{R^{(a,b)}}_{\mathrm{Col}}
 +\norm{(R^{(a,b)})^T}_{\mathrm{Col}}
 &\le2\sqrt n\,\norm{R^{(a,b)}}_F
 \le4Bn^{3/2}.
 \label{eq:csp-block-norms}
\end{align}
The first inequality uses
$\norm{R^{(a,b)}}_{\infty\to1}
\le n\norm{R^{(a,b)}}_{\mathrm{op}}$, as in
\Cref{subsec:residual}.

Applying \Cref{lem:restriction} with $p=s/n$ and substituting
\eqref{eq:csp-block-diagonal} and \eqref{eq:csp-block-norms}, we obtain
\begin{align*}
 \frac{\E[\norm{P_{\mathrm{var}}R^{(a,b)}P_{\mathrm{var}}}_{\infty\to1}]}{s^2}
 &\le C_{\mathrm{RV}}\left(
 \tau+\frac{2B^2}{\tau n}
       +\frac{2B^2}{\tau s}
       +\frac{4B}{\sqrt s}\right)\\
 &\le C_{\mathrm{RV}}\left(
 \tau+\frac{4B^2}{\tau s}
       +\frac{4B}{\sqrt s}\right),
\end{align*}
where the second inequality uses $s\le n$.
Summing over the four choices of $(a,b)$ in
\eqref{eq:csp-residual-blocks} proves the lemma because $B=16$ is
an absolute constant.
\end{proof}

\subsection{The sampling error for $L$}
If no eigenvalue is retained, then $L=0$ and the middle term of
\eqref{eq:csp-error-decomposition} vanishes. The bounds below hold
trivially in that case. We therefore assume that at least one
eigenvalue is retained. Let $j$ range over those eigenvalues and put
$$
 \lambda_j=\frac{\theta_j}{n},\qquad
 \phi_j(i,a)=\sqrt n\,u_j(i,a),\qquad
 y_{i,a}=(\lambda_j\phi_j(i,a))_j.
$$
Set $D=\diag(1/\lambda_j)$ and $f(w)=w^TDw$.
Here $|\lambda_j|>\tau$, so $\norm D_{\mathrm{op}}\le1/\tau$.
The spectral expansion gives, for every $z\in\mathcal X$,
\begin{align}
 f\left(\frac1n\sum_{i,a}z_{i,a}y_{i,a}\right)
 &=\sum_j\lambda_j
       \left(\frac1n\sum_{i,a}z_{i,a}\phi_j(i,a)\right)^2
 =\frac{z^TLz}{n^2},\notag\\
 f\left(\frac1s\sum_{i,a}\xi_i z_{i,a}y_{i,a}\right)
 &=\frac{z^TPLPz}{s^2}.
 \label{eq:csp-moment-objectives}
\end{align}
Thus the two retained maxima are $\max_K f$ and $\max_{K_Q}f$ on
the compact convex sets
$$
 K=\left\{\frac1n\sum_{i,a}z_{i,a}y_{i,a}:z\in\mathcal X\right\},
 \qquad
 K_Q=\left\{\frac1s\sum_{i,a}\xi_i z_{i,a}y_{i,a}:
                    z\in\mathcal X\right\}.
$$
Set $\Delta=d_H(K_Q,K)$ and
$t_y=\frac1n\sum_{i,a}\norm{y_{i,a}}^2$.
The next lemma is the counterpart of \Cref{lem:hausdorff-moments}.

\begin{lemma}[MAX-2CSP Hausdorff moments]\label{lem:csp-hausdorff}
For the sets above,
$$
 \E[\Delta]\le4\sqrt{t_y/s},\qquad
 \E[\Delta^2]\le\frac{17t_y}{s}.
$$
\end{lemma}
\begin{proof}

Recall from \Cref{eq:hausdorff-support} that for two nonempty compact and convex sets, their Hausdorff distance is exactly the largest difference of their support functions on unit vectors.

For $\norm v\le1$, put
$m_i(v)=\max_{a\in\{0,1\}}\ip v{y_{i,a}}$.
The support functions are
$$
 h_K(v)=\frac1n\sum_i m_i(v),\qquad
 h_{K_Q}(v)=\frac1s\sum_i\xi_i m_i(v).
$$
By \eqref{eq:hausdorff-support} and $p=s/n$,
\begin{equation}\label{eq:csp-hausdorff-selector}
 \Delta=\frac1s\sup_{\norm v\le1}
       \left|\sum_i(\xi_i-p)m_i(v)\right|.
\end{equation}

We use the ghost-sample symmetrization from the proof of
\Cref{lem:hausdorff-moments}. Let $\xi_i'$ be independent
Bernoulli$(p)$ selectors and let $\Xi_i$ be independent uniform
signs, independent of both samples. Since
$\E_{\xi'}[\xi_i']=p$ and
$(\xi_i-\xi_i')_i\overset{\mathrm d}{=}
(\Xi_i(\xi_i-\xi_i'))_i$, Jensen's inequality and the triangle
inequality give
\begin{align}
 \E[\Delta]
 &\le\frac1s\E_{\xi,\xi'}\left[
   \sup_{\norm v\le1}
   \left|\sum_i(\xi_i-\xi_i')m_i(v)\right|\right]\notag\\
 &\le\frac2s\E_{\xi,\Xi}\left[
   \sup_{\norm v\le1}
   \left|\sum_i\Xi_i\xi_i m_i(v)\right|\right].
 \label{eq:csp-hausdorff-symmetrization}
\end{align}

Define $c_i=(y_{i,0}+y_{i,1})/2$ and
$d_i=(y_{i,0}-y_{i,1})/2$. Then
\begin{align*}
 m_i(v)
 &=\max\{\ip v{y_{i,0}},\ip v{y_{i,1}}\}\\
 &=\frac12\bigl(\ip v{y_{i,0}}+\ip v{y_{i,1}}\bigr)
   +\frac12\bigl|\ip v{y_{i,0}}-\ip v{y_{i,1}}\bigr|\\
 &=\ip v{c_i}+|\ip v{d_i}|.
\end{align*}
Conditional on the selectors, the identity above and the triangle
inequality give
\begin{align*}
 &\E_{\Xi}\left[
   \sup_{\norm v\le1}
   \left|\sum_i\Xi_i\xi_i m_i(v)\right|\right]\\
 \le&\E_{\Xi}\left[
   \sup_{\norm v\le1}
   \left|\sum_i\Xi_i\xi_i\ip v{c_i}\right|\right]
 +\E_{\Xi}\left[
   \sup_{\norm v\le1}
   \left|\sum_i\Xi_i\xi_i|\ip v{d_i}|\right|\right]\\
 \le&\E_{\Xi}\left[\left\|\sum_i\Xi_i\xi_i c_i\right\|\right]
 +2\E_{\Xi}\left[
   \sup_{\norm v\le1}\sum_i\Xi_i\xi_i\ip v{d_i}\right]\\
 =&\E_{\Xi}\left[\left\|\sum_i\Xi_i\xi_i c_i\right\|\right]
 +2\E_{\Xi}\left[\left\|\sum_i\Xi_i\xi_i d_i\right\|\right].
\end{align*}
The $c_i$ term uses $\sup_{\norm v\le1}|\ip vz|=\norm z$.
For the $d_i$ term, the second inequality applies
\Cref{lem:rademacher-contraction} as in
\eqref{eq:hausdorff-contraction}.
For the $c_i$ term, independence and the mean-zero signs give
\begin{align*}
 \E_{\xi,\Xi}\left[\left\|\sum_i\Xi_i\xi_i c_i\right\|^2\right]
 &=\sum_i\E_\xi[\xi_i^2]\norm{c_i}^2
 +2\sum_{i<j}\E_{\xi,\Xi}[\Xi_i\Xi_j\xi_i\xi_j]\ip{c_i}{c_j}\\
 &=p\sum_i\norm{c_i}^2.
\end{align*}
Jensen's inequality then yields
\begin{align*}
 \E_{\xi,\Xi}\left[\left\|\sum_i\Xi_i\xi_i c_i\right\|\right]
 &\le\left(\E_{\xi,\Xi}\left[
     \left\|\sum_i\Xi_i\xi_i c_i\right\|^2\right]\right)^{1/2}
 =\left(p\sum_i\norm{c_i}^2\right)^{1/2},\\
 \E_{\xi,\Xi}\left[\left\|\sum_i\Xi_i\xi_i d_i\right\|\right]
 &\le\left(p\sum_i\norm{d_i}^2\right)^{1/2}.
\end{align*}
Since
$\sum_i(\norm{c_i}^2+\norm{d_i}^2)=nt_y/2$,
Cauchy--Schwarz and \eqref{eq:csp-hausdorff-symmetrization} give
\begin{align}
 \E[\Delta]
 &\le\frac2s\left[
     \left(p\sum_i\norm{c_i}^2\right)^{1/2}
    +2\left(p\sum_i\norm{d_i}^2\right)^{1/2}\right]\notag\\
 &\le\sqrt{\frac{10t_y}{s}}
 \le4\sqrt{\frac{t_y}{s}}.
 \label{eq:csp-hausdorff-first-moment}
\end{align}

For the second moment, let $\Delta^{(i)}$ be obtained by replacing
$\xi_i$ by its independent copy $\xi_i'$. The reverse triangle
inequality for the supremum in
\eqref{eq:csp-hausdorff-selector} gives
\begin{align*}
 |\Delta-\Delta^{(i)}|
 &\le\frac{|\xi_i-\xi_i'|}{s}
       \sup_{\norm v\le1}|m_i(v)|\\
 &\le\frac{|\xi_i-\xi_i'|}{s}
       \left(\norm{y_{i,0}}^2+\norm{y_{i,1}}^2\right)^{1/2}.
\end{align*}
Since $\E[(\xi_i-\xi_i')^2]=2p(1-p)$, the
Efron--Stein inequality (\Cref{prop:efron-stein}) gives
\begin{align*}
 \Var\Delta
 &\le\frac12\sum_i\E[(\Delta-\Delta^{(i)})^2]\\
 &\le\frac{p(1-p)}{s^2}
       \sum_i\left(\norm{y_{i,0}}^2+\norm{y_{i,1}}^2\right)
 =\frac{(1-p)t_y}{s}\le\frac{t_y}{s}.
\end{align*}
Combining this with \eqref{eq:csp-hausdorff-first-moment},
$$
 \E[\Delta^2]=\Var\Delta+(\E[\Delta])^2
 \le\frac{17t_y}{s}.
$$
\end{proof}

\begin{lemma}[Sampling the retained part]\label{lem:csp-retained}
$$
 \E\left[\left|
 \max_{z\in\mathcal X}\frac{z^TPLPz}{s^2}
 -\max_{z\in\mathcal X}\frac{z^TLz}{n^2}
 \right|\right]
 \le\frac{16B}{\sqrt s}+\frac{68B^2}{\tau s}.
$$
\end{lemma}
\begin{proof}
If $L=0$, both maxima vanish. Otherwise, by
\eqref{eq:csp-moment-objectives}, we must bound
$|\max_{K_Q}f-\max_K f|$. The key gradient bound is the same as
in the MAX-CUT proof. For $w=\frac1n\sum_{i,a}z_{i,a}y_{i,a}\in K$,
$$
 (Dw)_j=\frac1n\sum_{i,a}z_{i,a}\phi_j(i,a)
        =\frac1{\sqrt n}\ip{z}{u_j}.
$$
The eigenvectors are orthonormal, and
$\norm{z}_2^2\le n$ for $z\in\mathcal X$. Thus
\begin{equation}\label{eq:csp-gradient}
 \norm{Dw}^2
 =\frac1n\sum_j\ip{z}{u_j}^2
 \le\frac{\norm z_2^2}{n}\le1,\qquad
 \norm{\nabla f(w)}=2\norm{Dw}\le2.
\end{equation}
Since $f(w)=w^TDw$, its exact second-order expansion at $w\in K$
gives, for any $w'$ within distance $\Delta$ of $w$,
$$
 |f(w')-f(w)|
 =\left|2\ip{Dw}{w'-w}+(w'-w)^TD(w'-w)\right|
 \le2\Delta+\frac{\Delta^2}{\tau}.
$$
Here we used the gradient bound in \eqref{eq:csp-gradient} and
$\norm D_{\mathrm{op}}\le1/\tau$.
Choose $w_*\in K$ with $f(w_*)=\max_K f$ and $w'\in K_Q$ with
$\norm{w'-w_*}\le\Delta$. Applying the preceding bound at $w_*$ gives
$$
 \max_{K_Q}f\ge f(w')
 \ge f(w_*)-2\Delta-\frac{\Delta^2}{\tau}
 =\max_K f-2\Delta-\frac{\Delta^2}{\tau}.
$$
Conversely, choose $w'_*\in K_Q$ with $f(w'_*)=\max_{K_Q}f$ and
$w\in K$ with $\norm{w'_*-w}\le\Delta$. Applying the bound at $w$
gives
$$
 \max_{K_Q}f=f(w'_*)
 \le f(w)+2\Delta+\frac{\Delta^2}{\tau}
 \le\max_K f+2\Delta+\frac{\Delta^2}{\tau}.
$$
Combining the two inequalities yields
\begin{equation}\label{eq:csp-maxima-perturbation}
 |\max_{K_Q}f-\max_K f|
 \le2\Delta+\frac{\Delta^2}{\tau}.
\end{equation}

Orthonormality of the $u_j$ and \eqref{eq:csp-matrix-energy} give
\begin{align}
 t_y=\frac1n\sum_{i,a}\norm{y_{i,a}}^2
 &=\sum_j\lambda_j^2
   \left(\frac1n\sum_{i,a}\phi_j(i,a)^2\right)\notag\\
 &=\sum_j\lambda_j^2
 \le\frac{\norm M_F^2}{n^2}\le4B^2.
 \label{eq:csp-vector-energy}
\end{align}
Taking expectations in \eqref{eq:csp-maxima-perturbation} and
applying \Cref{lem:csp-hausdorff} and
\eqref{eq:csp-vector-energy} now gives
\begin{align*}
 \E[|\max_{K_Q}f-\max_K f|]
 &\le2\E[\Delta]+\frac{\E[\Delta^2]}{\tau}\\
 &\le8\sqrt{t_y/s}+\frac{17t_y}{\tau s}
 \le\frac{16B}{\sqrt s}+\frac{68B^2}{\tau s}.
\end{align*}
The identities in \eqref{eq:csp-moment-objectives} prove the claim.
\end{proof}

\subsection{Completing the bound}
\begin{proof}[Proof of \Cref{thm:csp-sampling}]
Take expectations in \eqref{eq:csp-error-decomposition} and apply
\eqref{eq:csp-full-residual} and
\Cref{lem:csp-residual,lem:csp-retained}. This gives
$$
 \E\left[\left|
 \frac{\mathrm{OPT}(F[Q])}{s^2}
 -\frac{\mathrm{OPT}(F)}{n^2}\right|\right]
 \le C_{\mathrm{res}}
  \left(\tau+\frac1{\tau s}+\frac1{\sqrt s}\right)
  +\frac{16B}{\sqrt s}+\frac{68B^2}{\tau s}+\tau.
$$
Set $\tau=1/\sqrt s\in(0,1]$. Since $B=16$ is fixed, the
right-hand side is at most $C_0/\sqrt s$ for an
absolute constant $C_0$.
\end{proof}

\begin{proof}[Proof of \Cref{thm:csp-upper}]
Let $s=\lceil C/\eps^2\rceil$ as in
\Cref{alg:csp-canonical}. If $n\le s$, querying the whole instance
uses $O(s^2)=O(1/\eps^4)$ pair queries and returns its exact optimum.
Assume $n>s$.

The algorithm can fail only if $|Q|>2s$, causing an early return,
or if its rescaled induced optimum has error greater than
$\eps n^2$. The variable selectors are independent Bernoulli$(s/n)$
variables, so $|Q|$ has mean $s$. Applying the Chernoff bound
(\Cref{prop:chernoff}) with deviation parameter $1$ gives
$$
 \Pr(|Q|>2s)\le e^{-s/3}\le e^{-C/3}.
$$
By \Cref{thm:csp-sampling} and Markov's inequality,
$$
 \Pr\left(\left|
 \frac{\mathrm{OPT}(F[Q])}{s^2}
 -\frac{\mathrm{OPT}(F)}{n^2}\right|>\eps\right)
 \le\frac{C_0}{\eps\sqrt s}
 \le\frac{C_0}{\sqrt C}.
$$
Choose the absolute constant $C$ so that
$e^{-C/3}+C_0/\sqrt C<1/3$.
The union bound proves the claimed success probability.

For the running time, enumerating all assignments to $F[Q]$ and calculating its maximum takes $2^{O(s)}=2^{O(1/\eps^2)}$ time.

\end{proof}

\end{document}